\documentclass[]{aa}  

\usepackage{graphicx}
\newcommand{\rJ}{$r_J\ $}
\newcommand{\rISCO}{$r_{isco}\ $}
\newcommand{\mdot}{$\dot{m}_{in}\ $}

\usepackage[final]{hyperref} 
\hypersetup{
	colorlinks=true,       
	linkcolor=blue,        
	citecolor=blue,        
	filecolor=magenta,     
	urlcolor=blue         
}

\usepackage{txfonts}
\usepackage[dvipsnames]{xcolor}
\usepackage{siunitx}
\usepackage{float}
\usepackage[normalem]{ ulem }
\usepackage{soul}
\usepackage{comment}
\begin{document} 

\title{Clues on jet behavior from simultaneous radio-X-ray fits of GX 339-4}

\author{S. Barnier\inst{1}
          \and
          P.-O. Petrucci\inst{1}
          \and
          J. Ferreira\inst{1}
          \and
          G. Marcel\inst{2,3}
          \and
          R. Belmont\inst{5}
          \and
          M. Clavel\inst{1}
          \and
          S. Corbel\inst{5,6}
          \and
          M. Coriat\inst{4}
          \and 
          M. Espinasse\inst{5}
          \and
          G. Henri\inst{1}
          \and
          J. Malzac\inst{4}
          \and
          J. Rodriguez\inst{5}
          }

\institute{Univ. Grenoble Alpes, CNRS, IPAG, 38000 Grenoble, France\\
              \email{samuel.barnier@univ-grenoble-alpes.fr}
         \and
             Villanova University, Department of Physics, Villanova, PA 19085, USA
         \and
             Institute of Astronomy, University of Cambridge, Madingley Road, Cambridge, CB3 OHA, United Kingdom 
         \and
             IRAP, Université de toulouse, CNRS, UPS, CNES , Toulouse, France
         \and 
             AIM, CEA, CNRS, Université de Paris, Université Paris-Saclay, F-91191 Gif-sur-Yvette, France
        \and 
            Station de Radioastronomie de Nançay, Observatoire de Paris, PSL Research University, CNRS, Univ. Orléans, 18330 Nançay, France
            }

\date{Received ... .., 2020; accepted --- --, 2020}

\abstract{Understanding the mechanisms of accretion-ejection during X-ray binaries' outbursts (XrB) has been a problem for several decades. For instance, it is still not clear yet what controls the spectral evolution of these objects from the hard to the soft states and then back to the hard states at the end of the outburst, tracing the well-known hysteresis cycle in the hardness-intensity diagram. Moreover, the link between the spectral states and the presence/absence of radio emission is still highly debated. In a series of papers, we developed a model composed of a truncated outer standard accretion disk (SAD, from the solution of Shakura and Sunyaev) and an inner jet emitting disk (JED). In this paradigm, the JED plays the role of the hot corona while simultaneously explaining the presence of a radio jet.
Our goal is to apply for the first time direct fitting procedures of the JED-SAD model to the hard states of four outbursts of GX 339-4 observed during the 2000-2010 decade by \textit{RXTE}, combined with simultaneous or quasi simultaneous \textit{ATCA} observations.
We built JED-SAD model tables usable in {\sc Xspec} as well as a reflection model table based on the {\sc Xillver} model of {\sc Xspec}. We apply our model to the 452 hard state observations obtained with \textit{RXTE}/PCA.
We were able to correctly fit the X-ray spectra and simultaneously reproduce the radio flux with an accuracy better than 15\%. We show that the functional dependency of the radio emission on the model parameters (mainly the accretion rate and the transition radius between the JED and the SAD) is similar between all the rising phases of the different outbursts of GX 339-4. But it is significantly different from the functional dependency obtained in the decaying phases. This result strongly suggests a change in the radiative and/or dynamical properties of the ejection between the beginning and the end of the outburst. We discuss possible scenarios that could explain these differences.

}

   \keywords{Black hole physics --
                X-rays: binaries --
                Accretion, accretion discs --
                ISM: jets and outflows 
               }

\maketitle

\section{Introduction}
  \label{introJEDSAD}

X-ray binaries (XrB) represent formidable laboratories to study the accretion-ejection processes around compact objects. Most of the time in a quiescent state, they can suddenly enter an outburst that can last a few months to a year, increasing their overall luminosity by several orders of magnitude. The X-ray emission is commonly believed to be produced by the inner regions of the accretion flow whereas the radio emission is considered as originated from relativistic jets. Simultaneously with the X-ray  spectral evolution along the outburst (from hard to soft states, e.g., \citealt{rem06,don07}), the radio emission switches from "jet-dominated" states during the hard X-ray states, at the beginning and the end of the outburst, to "jet-quenched" states during the soft X-ray states, in the central part of the outburst (e.g., \citealt{corbel2004origin,fender2004grs}). 
These outbursts are usually represented in the so-called hardness-intensity diagram (HID hereafter) where they follow a typical "q" shape (See e.g., \citealt{dun10}).

A physical understanding has yet to be found to explain the complete behavior of these outbursts even if a few points have reached consensus. For instance, the start of the outburst is believed to originate from disk instabilities in the outer regions of the accretion flow, driven by the ionization of hydrogen above a critical temperature (e.g., \citealt{ham98,fra02}). The nature of the soft state X-ray emission, peaking in the soft X-rays, is commonly attributed to the presence of an accretion disk, down to the Innermost Stable Circular Orbit (ISCO), and the standard accretion disk model (hereafter SAD, \citealt{shakura1973black}) seems to describe the observed radiative output reasonably well. The exact nature of the hard X-ray emitting region, the hot corona, is however less clear. Given its intense luminosity, it is expected to be located close to the black hole, where the release of gravitational power is the largest. The variability of the source is consistent with a very compact region (\citealt{de2017evolution}), but the exact geometry is still matter to debate. It could be located somewhere above the black hole (the so-called lamppost geometry, e.g., \citealt{matt1991iron,martocchia1996iron,miniutti2004light}).  
It can also partly cover the accretion disk (the patchy corona geometry, e.g., \citealt{haa97}). Or it can fill the inner part of the accretion flow, the accretion disk being present in the outer part of the flow (the corona-truncated disk geometry, e.g., \citealt{esin1997advection}). Of course, the hot corona is most probably a combination of all these geometries, or it could even evolve from one geometry to the other depending on the state of the source.

The dominant radiative process producing the hard X-rays is generally believed to be external Comptonization, meaning Comptonization of the external UV/soft X-ray photons produced by the accretion disk off the hot electrons present in the corona. These electrons are generally supposed to follow a relativistic thermal distribution to explain the presence of a high-energy cutoff, generally observed in the brightest hard states (see e.g., \citealt{fab17}, for a recent compilation). While the release of the gravitation power is undoubtedly the source of the corona heating, how this heating is transferred to the particles and how particles reach thermal equilibrium is still not understood. The magnetic field is expected to play a major role (e.g., \citealt{mer01}) but the details of the process are unknown. The correlation between X-rays (from the corona) and radio (from the jet) emission (e.g., \citealt{gallo2003universal,cor00,cor03,cori11,gallo2012assessing,cor13}) also indicates a strong link between accretion and ejection; supporting the presence of a magnetic field in the disk.

Models of the hot corona commonly used in the literature are generally oversimplified. The geometry is assumed to have a basic shape, for instance spherical, slab or even point-like. Its temperature and density are supposed to be uniform across the corona. External Comptonization is usually the unique radiative process taken into account and the spectral emission is often approximated by a cut-off power law (or similar) shape. Most of the time, no physically motivated configuration or comparison with numerical simulations of these simplified model is proposed.
More importantly, the jet emission and its impact on the accretion system is generally entirely ignored.\\

\begin{figure*}[t!]
            {\includegraphics[width=\textwidth]{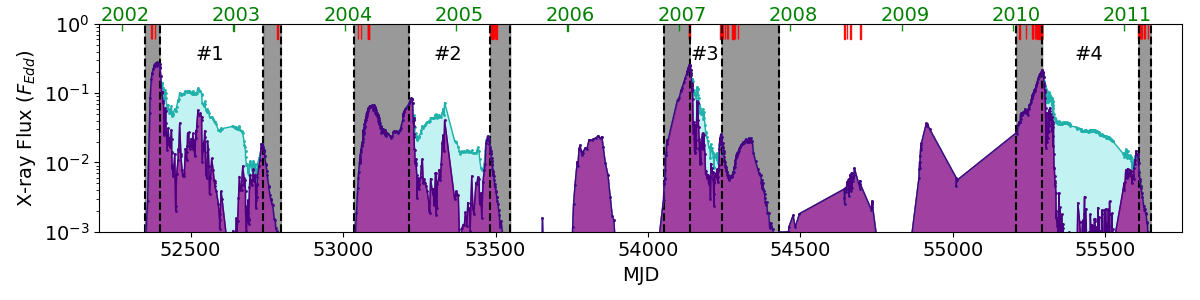}}
      \caption{GX 339-4 X-ray light curve in the 3-200 keV energy band of the 2000-2010 decade obtained with the \citet{clavel2016systematic} fits. The violet filled region shows the power-law unabsorbed flux while the cyan region represents the disk unabsorbed flux. Highlighted in grey, we show the selected spectra for this study : the rising and decaying hard states of the 4 outbursts (\#1, \#2, \#3 and \#4). At the top red lines represent the date where steady radio fluxes were observed at 9 GHz (from \citealt{cor13}).}
         \label{Fig:GX_lightcurve}
\end{figure*}

Magnetized accretion-ejection solutions that self-consistently treat both the accretion disk and the jets have been developed for more than 20 years \citep{fer95,fer97} and validated through numerical simulations since (e.g., \citealt{zanny2007},  \citealt{jacquemin2021magnetic}). In these works, the accretion disk is assumed to be threaded by a large-scale magnetic field. In the regions where the magnetization $\mu(r) = P_{mag}/P_{tot}$ is of the order of unity (with $P_{mag}$ the magnetic pressure and $P_{tot}$ the total pressure, sum of the thermal and radiation pressure), the magnetic hoop stress overcomes both the outflow pressure gradient and the centrifugal forces: self-confined, non-relativistic jets can be produced. In these conditions, the accretion disk is called a Jet Emitting Disk (JED). The effect of the jets on the disk structure can be tremendous since the jets torque can efficiently extract the disk angular momentum, significantly increasing the accretion speed. Consequently, for a given accretion rate, a JED has a much lower density in comparison to the standard accretion disk (e.g., \citealt{ferreira2006unified}). The parameter space for stationary JED solutions correspond to magnetization $\mu$ in the range [0.1,1], small ejection index $p <0.1$ defined by $\dot{m}(r)\propto r^{p}$ (with $\dot{m}$ the mass accretion rate measured at a given radius $r$), large sonic mach number $m_s=u_r/c_s$ in the range [1-3] (with $u_r$ the accretion speed and $c_s$ the local speed of sound) and jet power fraction $b=P_{jets}/P_{acc}$ between 0.1 to almost 1 for very thin JED \citep{fer97} -- where $P_{jets}$ is the power feeding the jets and $\displaystyle P_{acc}$ the total gravitational power released by the accretion flow within the JED. We note that for weak magnetization ($\mu \ll 0.1$), no collimation occurs and uncollimated winds are produced. The accretion disk structure is then not strongly different from the standard solution (e.g., \citealt{jac19}).\\

\citet{ferreira2006unified} proposed an hybrid disk paradigm to address the full accretion-ejection evolution of XrB in outbursts (see also \citealt{petrucci2008role}). The accretion disk is threaded by a vertical magnetic field and extends all the way down to the innermost circular orbit. The outer part of the flow has a low magnetization, resulting in an outer SAD. On the contrary, the inner part has an important mid-plane magnetization and the accretion flow has a JED structure. Such radial distribution of the magnetisation appears to be a natural outcome of the presence of large-scale magnetic fields in the accretion flow, the magnetic flux accumulating toward the center to produce a magnetized disk with a fast accretion timescale (\citealt{scepi2020magnetic}, \citealt{jacquemin2021magnetic}). 

\citet{marcel2018bunified} developed a two temperature plasma code to compute the Spectral Energy Distribution (SED) of any JED-SAD configuration. In addition to the parameter $\mu$, $p$, $m_s$ and $b$ that characterizes a JED solution, the output SED also depends on 
the transition radius \rJ (in unit of gravitational radius $\displaystyle R_G = GM/c^2$) between the JED and the SAD, as well as the mass accretion rate \mdot (in unit of Eddington accretion mass rate $\dot{M}_{Edd}=L_{Edd}/c^2$) reaching the ISCO. 

In our view, \rJ and $\dot{m}_{in}$ are expected to be physically linked through the evolution the magnetization across the accretion flow, dividing it in an inner strongly magnetized part (the JED) and an outer weakly magnetized one (the SAD). Now, this link, is far from being trivial to estimate and require global 3D MHD simulations, far out of the scope of our present modeling, to catch it. In the absence of any physical law that could be used as an input, we consider the parameters \rJ and $\dot{m}_{in}$ as independent parameters.

Varying \rJ and $\dot{m}_{in}$, \citet{marcel2018bunified} showed that the JED-SAD model was able to qualitatively reproduce the spectral evolution of an entire outburst. The JED radiative properties agree with the X-ray emission observed in compact objects \citep{pet10,marcel2018aunified,marcel2018bunified}, meaning that the JED can play the role of the hard X-ray emitting hot corona. 
At the beginning of the outburst, the disk is characterized by a low luminosity hard component that can be represented by a JED with large radial extend (\rJ $\gg$ 1), while the mass accretion rate \mdot is low (\mdot $<$ 0.1). During the rising part of the outburst, the luminosity increases, thus the mass accretion rate increases but \rJ is still several times the ISCO radius. While transitioning to the soft states, \rJ starts to decrease until the complete disappearance of the JED when $r_J \sim r_{ISCO}$. This coincides with the disappearance of the radio emission (no more JED implies no more jets). During the soft states, the disk is dominated by the thermal component of the SAD and \rJ stays equal to $r_{ISCO}$. Eventually, during the decaying phase of the outburst, a JED reappears when the system transitions back to the hard state, \mdot decreases, \rJ increases again and the system returns to the hybrid JED-SAD configuration. With the following decrease of the accretion rate, the XrB then fades to the quiescent state.\\

\citet{marcel2019unified} (hereafter M19) performed the first application of the JED-SAD model to real data by qualitatively reproducing the spectral evolution of GX 339-4 during the 2010 outburst observed by RXTE. A similar study has been recently extended to three others outbursts of GX 339-4 (\citealt{mar20}, hereafter M20). These authors did not directly fit the data given the large number of observations as well as the lack of a consistent reflection model component. Instead, they produced a large grid of spectra for a set of parameters ($r_J$, $\dot{m}_{in}$) and fitted each simulated spectrum with a disk + power law model. This provided spectral characteristics (disk flux, power law luminosity fraction, X-ray spectral index) that were then compared to the best fit results obtained by fitting the RXTE/PCA data with a similar disk + power law model \citep{clavel2016systematic}. This procedure allowed us to derive the qualitative evolution of \rJ and \mdot that reproduces JED-SAD spectra with the closest spectral characteristics to the observed ones.\\ 
 
An important output of the JED-SAD model is the estimate of the jets power in a consistent way with the JED structure. The radio signature produced by this jet is less straightforward to estimate however, since it depends on the detailed treatment of the jet particle emission all along the jet. Following \citet{heinz2003} (hereafter HS03), M19 proposed an expression for the radio flux produced at a radio frequency $\nu_R$ by a jet launched from a JED characterized by \rJ and \mdot:
\begin{equation}
    F_{R} = \Tilde{f}_R \dot{m}_{in}^{17/12}r_{isco}  \, (r_J-r_{isco})^{5/6}\frac{F_{Edd}}{\nu_R}
    \label{eq:Fradio0}
\end{equation}
where $\Tilde{f}_R$ is a scaling factor and $\displaystyle F_{Edd}=L_{Edd}/4\pi d^2$ the Eddington flux. Equation (\ref{eq:Fradio0}) has the same dependency with the accretion rate than in the self-similar approach of HS03 but there is an additional multiplicative term 
$(r_J-r_{isco})^{5/6}$ that reflects the necessarily finite radial dimension of the jet  due to the finite radial dimension of the JED (see discussion in M19).
By trying to simultaneously reproduce the X-ray and radio emission, M19 were able to put constraints on \mdot, \rJ and $\Tilde{f}_R$ in the case of GX 339-4, assuming a constant $\Tilde{f}_R$ for all the outbursts. \footnote{However a better result was obtained when using a constant, but different, $\Tilde{f}_R$ for the rising and decaying phase with $\Tilde{f}_R^{rise}>\Tilde{f}_R^{decay}$ (G. Marcel, private communication).}\\ 

The present paper aims to make a step forward in the comparison of the JED-SAD model to real data through a direct fitting procedure of simultaneous radio and X-ray data of an X-ray binary. The improvements compared to the previous works are two-fold. First we add a consistent reflection component in the model and second we obtain more reliable and precise constrains on our model parameters (i.e., \rJ and $\dot{m}_{in}$). To do so, we develop the required tools to apply our JED-SAD model to standard fitting software (like {\sc{xspec}}, \citealt{1996ASPC..101...17A}).

We focus in this paper on the simultaneous radio-X-ray  coverage of the XrB GX 339-4 during the lifetime of the RXTE satellite (1995-2012). This corresponds to four major outbursts starting in 2002, 2004, 2007 and 2010. The data selection is discussed more precisely in Sect. \ref{sec:data}. Our fitting procedure and first fit results, using Eq. (\ref{eq:Fradio0}) for the radio emission, are discussed in Sect. \ref{sec:norad}. These results suggest however a different functional dependency of the radio emission with \rJ and \mdot compared to Eq. (\ref{eq:Fradio0}). A deeper analysis of the radio behavior is then performed in Sect. \ref{sec:rad} and supports two different functional behaviors of the radio emission between the beginning and the end of the outburst. The implication of these results are discussed in Sect. \ref{Sec:discuss} before concluding in Sect. \ref{sec:conclusion}.

\section{Data selection} \label{sec:data}
To test our JED-SAD paradigm we focus on simultaneous or quasi simultaneous radio/X-ray observations of GX 339-4. We only use "pure" hard states (i.e., those at the very right part of the HID), either in the rising or decaying phase, and do not include the transition phases of the outburst even when radio emission is detected (during the so-called Hard Intermediate state, HIS). The reasons for this choice are twofold. First, the radio flux is smoothly evolving during pure hard-states, a signature of stationary processes hopefully easier to catch. Conversely, an important radio variability is observed during the transition phases, especially during the hard-to-soft transition. Second, during the transition states, the hard-tail component progressively appears. As this component is not well understood and is not self-consistently included in the JED-SAD model, we do not select the transition states.

We selected X-ray spectra from the RXTE-PCA archive of GX 339-4 during the 2000-2010 decade\footnote{In order to have a uniform data analysis, we do not include the data from the RXTE/HEXTE instrument since there were not always usable (e.g., in the case of low flux observations or after March 2010 when it definitely stops observing).}. The data processing is detailed in \citet{clavel2016systematic}. Since the instrumental background was generally found to be of the order of, or larger than, the source emission above 25 keV, we limited our spectral analysis to the 3-25 keV energy range of the PCA instrument. 
We plotted the 2000-2010 PCA X-ray light curve of GX 339-4 in Fig. \ref{Fig:GX_lightcurve}. During this period, GX 339-4 undergoes four complete outbursts, in 2002, 2004, 2007 and 2010, hereafter outbursts \#1, \#2, \#3 and \#4. The hard-only or "failed" outbursts of 2006 and 2008 were not selected for this study\footnote{"Failed" outbursts only present hard states and no transition to the soft states before going back to quiescence.} since they may be intrinsically different from the ones accomplishing an entire HID. We follow \citet{clavel2016systematic} for the definition of the hard state periods of each outburst. We report in Table \ref{tab:mjd_hard} the corresponding starting and ending Modified Julian Dates (MJD) of both the rising and the decaying hard state phases.

   \begin{table}[t]
      \caption[]{Hard state periods of the 4 outbursts and number of selected observations.}
         \label{tab:mjd_hard}
         \centering
         \begin{tabular}{|c|c|c|c|c|}
            \hline
             & Rise$^{(a)}$ & Decay$^{(b)}$ & X-ray$^{(c)}$ & Radio$^{(d)}$\\[0.2ex]
            \hline
            \hline
            \#1 & 52345-52399 & 52739-52797 & 49 & 4 (3/1) \\
            \#2 & 53036-53219 & 53482-53549 & 177 & 16 (7/9)\\
            \#3 & 54051-54137 & 54241-54429 & 146 & 13 (2/11) \\
            \#4 & 55208-55293 & 55609-55640 & 80 & 24 (16/8) \\
            \hline
         \end{tabular}
        \tablefoot{  Hard state periods of the 4 outbursts as defined by \citet{clavel2016systematic}. $^{(a)}$ MJD of the rising phase of each outburst. $^{(b)}$ MJD of the decaying phase of each outburst. $^{(c)}$ Number of X-ray observations covering each outburst. $^{(d)}$ Number of radio observations covering each outburst, we specify the number of rising phase observations or decaying phase observations using the notation: (rising / decaying).}
    \end{table}

In radio, we used the 9 GHz fluxes obtained with the Australia Telescope Compact Array (ATCA) and discussed in \citet{cor13}\footnote{Before 2009 the radio band was 128 MHz wide and centered at 8.64 GHz. After 2009 it was 2 GHz wide and centered at 9 GHz.}. Compared to the X-ray observations, the radio survey is quite sparse (see Fig. \ref{Fig:GX_lightcurve}), so we selected only the radio fluxes close to X-ray pointings by less than one day (which we call quasi-simultaneous radio/X-ray observations).  \\

This selection corresponds to a total of 452 hard X-ray spectra and 57 radio fluxes distributed among the four outbursts. Outburst \#4 is the one with the best X-ray and radio coverage, with about 80 X-ray spectra and 24 radio measurements well distributed along the outburst.
Thanks to this large radio-coverage we choose to linearly interpolate the radio light curve to estimate the radio fluxes for each of the 80 X-ray spectra of this outburst. This is supported by the smooth evolution of the radio light-curve during the pure hard states. The resulting interpolation is plotted in Fig. \ref{fig:Radio_interp} in appendix \ref{sec:a_interp}. Such an interpolation was not possible for the other outbursts due to the too small number of radio pointings.

\section{X-ray and radio fits} \label{sec:norad}

\subsection{Methodology}

Similarly to M19, we assume a distance $d=8\,\mathrm{kpc}$ for GX 339-4 (\citealt{hynes2004distance,parker2016nustar}) and a black hole mass $M=5.8\, \mathrm{M_\odot}$ (\citealt{hynes2003dynamical,parker2016nustar}). The innermost stable circular orbit \rISCO is assumed equal to $2$ in $\mathrm{R_g}$ units. This is equivalent to a black hole spin of 0.94 (\citealt{miller2008initial,garcia2015x}). Finally, for the Galactic hydrogen column density we use $0.6\times 10^{22}\, \mathrm{cm^{-2}}$ (\citealt{zdziarski2004gx,bel2011overview}).

Concerning the JED-SAD model, the two parameters left free to vary during the fitting procedure are \rJ and $\dot{m}_{in}$.
All the other parameters of the JED-SAD (see Sect. \ref{introJEDSAD}) are set to the same values than in M19, that is $b=0.3$, $m_s=1.5$ 
and $p=0.01$\footnote{In the precedent papers (\citealt{marcel2018aunified,marcel2018bunified,marcel2019unified}), this parameter was called $\xi$. However to avoid the confusion with the ionization parameter of the reflection component, we introduce the notation \textit{p}.} \\

We create {\sc{xspec}} model tables for the JED and the SAD components separately with 40 values of \mdot and 25 values of \rJ logarithmicaly distributed in the range [0.001,10] and [1,300] respectively. We also produce a reflection table.
For that purpose, we use the {\sc{xillver}} reflection model  (\citealt{garcia2013x}). For each couple ($r_J$, $\dot{m}_{in}$) of the JED table, we fit the corresponding JED spectrum with a cut-off power law model. This fit provides a spectral index and a high-energy cutoff that we inject in the {\sc{xillver}} table to produce different reflection spectra for different values of the disk ionization $log(\xi)$ and iron abundance $A(Fe)$ (in solar unit). The disk inclination is set to 30$^{\circ}$, an inclination consistent with the one expected for GX 339-4 (\citealt{parker2016nustar}). 
The resulting table thus possesses five different parameters for each spectrum: the 2 JED-SAD parameters ($r_J$, $\dot{m}_{in}$), the three reflection parameters $log(\xi)$, $A(Fe)$ and the reflection normalisation.\\

We then use an automatic fitting procedure using the pyxspec library (a python interface to {\sc{xspec}}). We fit the X-ray spectra with the following {\sc{xspec}} model : {\texttt{tbabs $\ast$ (atable(JEDtable) + atable(SADtable) + kdblur$\ast$atable(Refltable))}} 

Where \texttt{JEDtable}, \texttt{SADtable} and \texttt{Refltable} are the {\sc{xspec}} tables for the JED, SAD and reflection spectra respectively, and {\sc{kdblur}} a convolution model of {\sc{xspec}} to take into account the relativistic effects from the accretion disk around a rotating black hole (according to the original calculations by \citealt{Laor1991}). The parameters \rJ and \mdot are tied together between each table and we set the inner radius of {\sc{kdblur}} to the inner radius of the SAD (i.e., $r_J$). In {\sc{kdblur}}, we freeze the index of the disk emissivity to 3 (its default value), the outer disk radius to 400 $\mathrm{R_g}$ and the inclination to 30$^{\circ}$.

\begin{figure*}[t!]
   \resizebox{0.85\hsize}{!}
            {\includegraphics[scale=1.5]{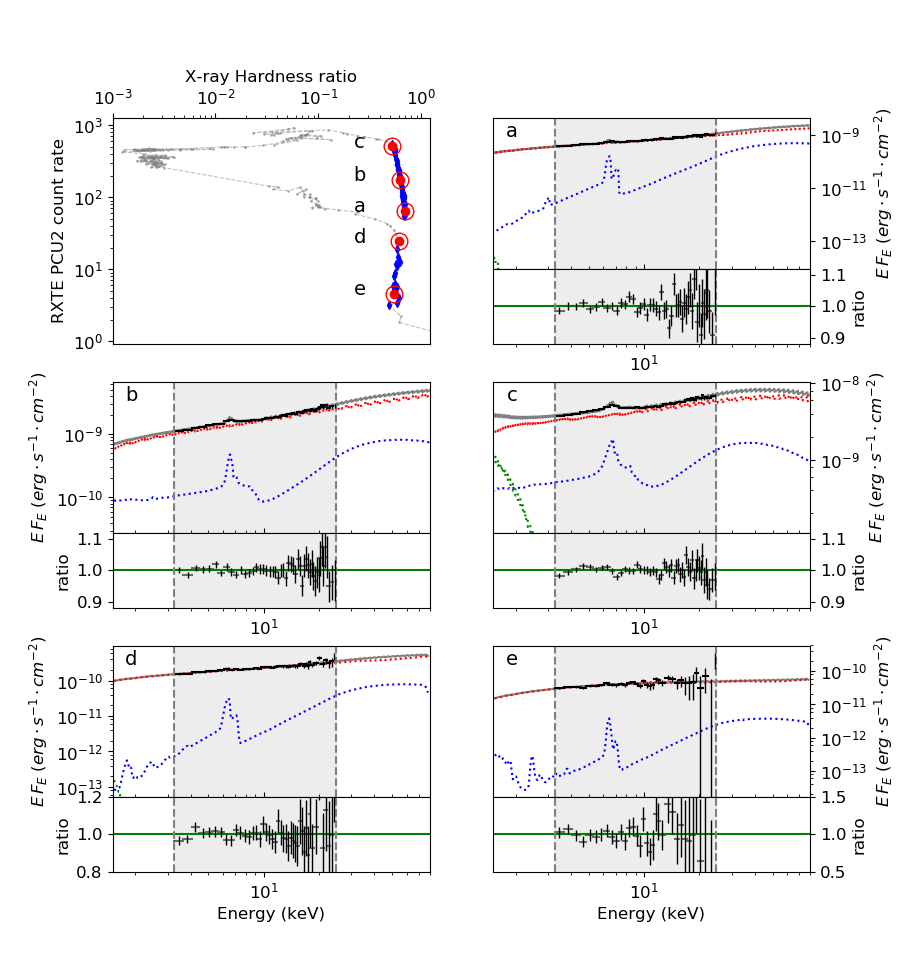}}       
            \centering

       \caption{Best fits of some observations of outburst \#4. In the \textbf{top left}, the Hardness Intensity Diagram of outburst \#4. The blue diamond show the hard state used for this outburst. The 5 red points are the 5 observations plotted in the different figures from a) to e). \textbf{a)-e):} Best fit spectra and data/model ratio for the 5 observations indicated in red in the HID. The grey region shows the PCA energy range used for the fit. The data are in black and the best fit model in grey, the JED spectrum is in red, the SAD spectrum in green and the reflection component in blue. The best fit parameters for each observation are reported Table \ref{tab:Fit_results}. }
    \label{Fig:showing_fits}
   \end{figure*}

\begin{table*}[]
    \caption{Fitting parameters of the 5 observations presented in Fig. \ref{Fig:showing_fits}}
    \centering
       \begin{tabular}{|c|cc|cc|cc|}
        \hline 
          Observations & MJD $^{(a)}$ & $\chi^{2}/DoF$ $^{(b)}$ & \rJ$^{(c)}$ & \mdot$^{(d)}$ & log($\xi$) $^{(e)}$ & \textit{N} $^{(f)}$ \\\hline
        \textit{a}  & 55214.089 & 42/45 & $44.0^{+2.5}_{-4.0}$ & $0.87^{+0.01}_{-0.03}$ & $<2.0$ & $9.9^{+0.2}_{-9.4}\times10^{-4}$  \\ 
        \textit{b}  & 55260.445 & 32/45 & $35.7^{+2.8}_{-2.1}$ & $1.25^{+0.03}_{-0.01}$ & $3.08^{+0.03}_{-0.02}$ & $1.3^{+0.1}_{-0.2}\times10^{-6}$\\ 
        \textit{c}  & 55292.779 & 59/45 & $14.3^{+0.6}_{-0.6}$ & $2.31^{+0.02}_{-0.02}$ & $3.22^{+0.10}_{-0.06}$ & $1.5^{+0.2}_{-0.2}\times10^{-6}$\\
        \textit{d}  & 55609.839 & 22/40 & $27.2^{+5.8}_{-4.5}$ & $0.37^{+0.04}_{-0.03}$ & $< 4.5$ & $< 2.0\times10^{-4}$\\
        \textit{e}  & 55634.085 & 23/31 & $ > 57$ & $7.4^{+0.5}_{-1.0}\times10^{-2}$ & $<4.6$ & $< 2.8\times10^{-4}$\\ \hline
       \end{tabular}
       \tablefoot{$^{(a)}$ MJD of the observations. $^{(b)}$ $\chi^2$ statistics of the fit and the number of degrees of freedom (DoF). $^{(c)}$ Transition radius $r_J$ in $R_G$. $^{(d)}$ Mass accretion rate $\dot{m}_{in}$ in $\dot{M}_{Edd}$. $^{(e)}$ Disk ionization $\xi$ from the reflection model. $^{(f)}$ Reflection normalization \textit{N}, units of the  {\sc{xillver}} reflection model.}
    \label{tab:Fit_results}
\end{table*}

\begin{figure*}
   \resizebox{1\hsize}{!}
            {\includegraphics[scale=1.5]{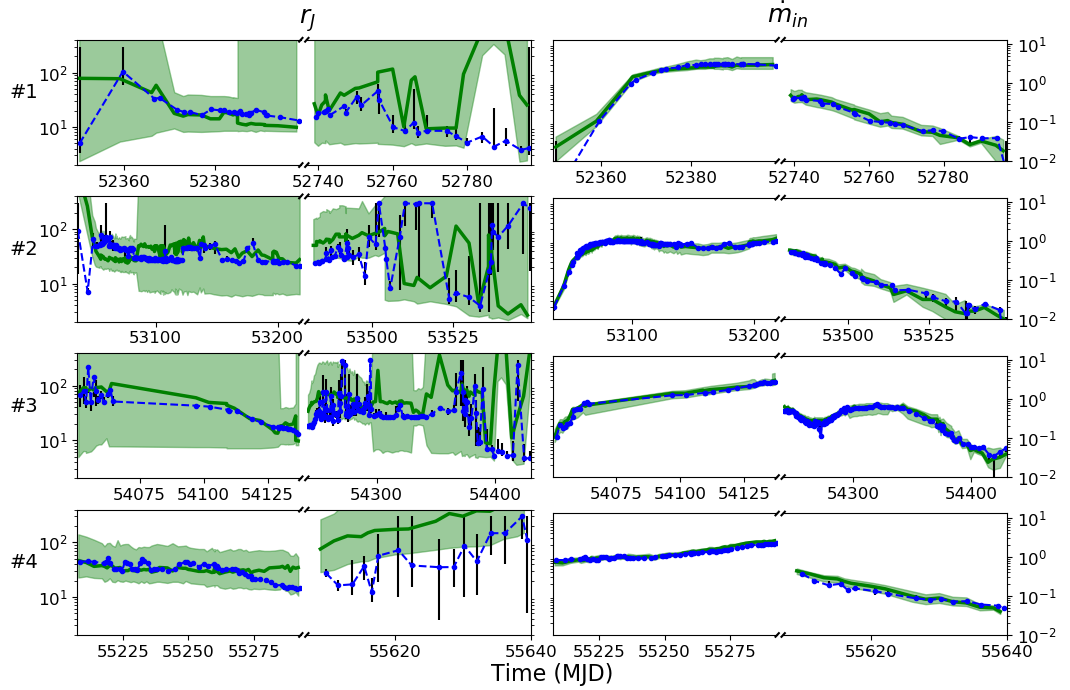}}
            
      \caption{Results of the fitting procedure. On the left side, the transition radius \rJ (from $r_{isco}$ to 300) between the JED and the SAD. On the right side, the mass accretion rate $\dot{m}_{in}$. Each side is divided vertically between the 4 outbursts and horizontally between the rising and decaying phase of each outburst. The green solid line represent the results from M19 and M20, and the green region where their minimization function varies by less than 10\% with respect to its minimum. The blue dashed line shows the results of the fitting procedure and the black vertical bar the associated 90\% confidence range. The decaying phase of outburst \#4 is the subject of appendix \ref{sec:b_decay2010}.}
         \label{Fig:Fits_results_norad}
         
\end{figure*}

\subsection{X-ray fits} \label{sec:X-rayfits}

Using the fitting procedure described above, we obtain the best fit values for $r_J$, $\dot{m}_{in}$ for each X-ray observation in our sample. The iron abundance clustered around 7 times the solar abundance, in agreement with similar spectral analysis of GX 339-4 (e.g., \citealt{garcia2015x,furst2015,parker2016nustar,wang2018evolution}) and we set it to this value in the following\footnote{Such a high iron abundance could be a consequence of the {\sc{xillver}} reflection model used. A new version of this model, with higher disk density, gives iron abundance closer to solar values (e.g., \citealt{tomsick2018,jiang2019}).}. As examples, we report in Fig. \ref{Fig:showing_fits} a few of our X-ray best fits obtained for different observations distributed in the hard X-ray states of outburst \#4. On the top left of this figure we represent the HID as well as the hard states (blue diamond) that we fit. We also highlight the five observations whose spectral fits are presented in the other panels of the figure. These panels present at the top the best fit model,
the grey highlighted zone represents the PCA energy range fitted. The black crosses show the PCA data. At the bottom of each panel, we present the ratio between the data and the model. The fits parameters for each of these observations can be found in the table at the bottom of Fig. \ref{Fig:showing_fits}.

During the rising phase (observations \textit{a}, \textit{b} and \textit{c}), the high-energy cutoff slowly appears in the model as we rise in luminosity.\footnote{Even though the high-energy cutoff is not visible in the energy range we fit, the JED-SAD parameters we obtain predict a decrease of the high-energy cutoff during the rising phase, similarly to what is observed (\citealt{motta2009evolution,droulans2010variability}).} 
At the same time, the iron line is changing shape under the influence of both the evolution of the disk ionization parameter and the black hole gravity as the transition radius $r_J$ decreases (general relativity effects). During the decaying phase (observations \textit{d} and \textit{e}), as the luminosity decreases, the standard accretion disk component disappears with the increase of $r_J$. \\

The evolutions of \rJ and \mdot for our entire data sample are reported in Fig. \ref{Fig:Fits_results_norad}, the left panel showing the light curves of \rJ and the right panel the ones of $\dot{m}_{in}$. We have subdivided each panel in two, showing the rising phase first and then the decaying phase. The large green region represents the area where the minimization function used by M19 to constrain \rJ and \mdot varies by less than 10\% with respect to its minimum. The blue points with black error bars and connected by dashed lines represent the results of this paper obtained by fitting the X-ray spectra in {\sc{xspec}}.

Clearly, we obtain much tighter constraints compared to M19, especially for \rJ during the rising phase. There are two reasons for this: first, M19 did not directly fit the data, their main objectives being to qualitatively reproduce the outburst spectral and flux evolution. Secondly M19 did not use a $\chi^2$ statistics to constrain their parameters, the $\chi^2$ statistics being not well-adapted to their methodology.

Nevertheless, the constraints obtained with our fitting procedure are almost always embedded within the green area obtained by M19, showing the good agreement between the two approaches. This is noticeably the case for \mdot which is well constrained in both methods and in very good agreement with each other.  Interestingly our values for \rJ are apparently better constrained in the rising phase of the outbursts, its behavior being more erratic and with larger error bars in the decaying phase. This could be a natural effect of the decrease of the data statistics when the flux decreases but this trend is not observed on $\dot{m}_{in}$. This instead suggests that our JED-SAD spectra are less dependent on \rJ at low accretion rate. 

It should be noted that the $\chi^2$ space is not always following a Gaussian shape (see Fig. \ref{fig:steppar_rj}) and the error bars should not be taken as $\sigma$ errors but instead as lower and upper limits with a 90\% confidence. Thus, none of the "pure" hard states are consistent with $r_{isco}$, and a JED is always required in the fit.

The evolution of \rJ during the decaying phase of outburst \#4 is the subject of appendix \ref{sec:b_decay2010} where we detail how we obtained the presented values of \rJ using a maximum likely-hood method. When a small $r_J$ solution was found in the automatic procedure ($r_J<10$) we checked the parameter space for a statistically equivalent solution at bigger value of $r_J$. Whenever such a solution exists we selected it (see appendix \ref{sec:b_decay2010}). The motivations for this choice are twofold: higher value of \rJ are observed in the decaying phase of the other outbursts (see Fig. \ref{Fig:Fits_results_norad} \#2 and \#3) and the resulting increase in \rJ when going to quiescence is consistent with the JED-SAD dynamical picture.

\begin{figure*}[h!]
   \centering
            { \includegraphics[width=\textwidth]{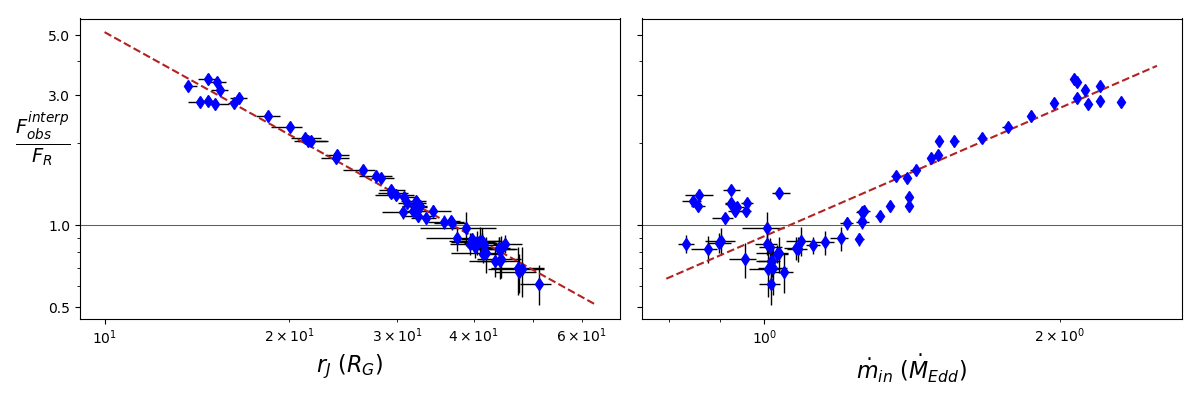}}
            
      \caption{Ratio between the observed radio fluxes and the results of Eq. (\ref{eq:Fradio0}) (using $\tilde{f}=1.5\times 10^{10}$) in function of \rJ (left) and \mdot (right) for outburst \#4. In the case of X-ray observations without simultaneous radio measurement, the radio flux was interpolated from the radio light curve (see Fig. \ref{fig:Radio_interp}). The dashed line show the best fit power law:  $\frac{F_{obs}}{F_R}\propto$ \rJ$^{-1.25}$ (left) and $\frac{F_{obs}}{F_R} \propto$ \mdot$^{1.56}$ (right).}
    \label{fig:ftilderj}
\end{figure*}

\subsection{Taking into account the radio emission}
We now reproduce the radio fluxes using Eq. (\ref{eq:Fradio0}) to model the radio emission. Since the radio survey is generally quite sparse we first concentrate on the rising phase of outburst \#4, where the radio coverage is sufficiently dense to interpolate the radio fluxes for all X-ray spectra (see Fig. \ref{fig:Radio_interp}).
We use the results of the X-ray fits (see previous subsection) and set the parameters \rJ and \mdot to the best fit values. Then we compute the radio flux $F_R$ with Eq. (\ref{eq:Fradio0}) using $\tilde{f}=1.5\times 10^{-10}$, the value used in M19. In Fig. \ref{fig:ftilderj} we have plotted the ratio between the observed (and interpolated) radio flux $F_{obs}$ and the expected radio flux $F_R$ from Eq. (\ref{eq:Fradio0}) as function of \rJ and $\dot{m}_{in}$. A clear anticorrelation is observed $\displaystyle\frac{F_{obs}}{F_R}\propto r_J^{\alpha}$ with $\alpha\sim -1.25$. Similarly, $\displaystyle\frac{F_{obs}}{F_R}$ is correlated with $\dot{m}_{in}$, with a power $\beta\sim 1.56$. In conclusion here our fitting procedure suggests a functional dependency of the radio emission at least on \rJ and/or \mdot that is not taken into account correctly when using Eq. (\ref{eq:Fradio0}). This is deeply studied in Sect. \ref{sec:rad}.

\section{Functional dependency of the radio emission} \label{sec:rad}
In the JED-SAD paradigm, the evolution of the X-ray spectrum (hardness, energy cutoff and flux) is described through the changes of two parameters, \rJ and $\dot{m}_{in}$, controlling the balance between the power released through advection and radiation. In a similar way, we will use both of these parameters to describe the radio flux.

We thus assume in this section a more general expression for the radio flux:
\begin{equation}
    F_{R} = \Tilde{f}^{*} \,r_J^\alpha  \,\dot{m}_{in}^{\beta} \, \bigg(1-\frac{r_{isco}}{r_J}\bigg)^{5/6}\,\frac{F_{Edd}}{\nu_R}. 
    \label{eq:Fradio1}    
\end{equation}
This expression is relatively similar to Eq. (\ref{eq:Fradio0}), but the indexes of the dependency on both \mdot and \rJ are now free parameters. This new expression allows us to put all the dependency on $r_J$ and $\dot{m}_{in}$ in the parameters $\alpha$ and $\beta$, $\Tilde{f}^{*}$ acting then as a true constant in this respect. The term $(r_J-r_{isco})^{5/6}$ linked to the radial extension of the jet is re-expressed to isolate the dominant power dependency with $r_J$ in $\alpha$. We look for a unique triplet ($\Tilde{f}^{*}$, $\alpha$, $\beta$) that could reproduce the whole radio data set. 

\subsection{Rising phase of the 2010 outburst}

\begin{figure}[!]
    {\includegraphics[width=\columnwidth]{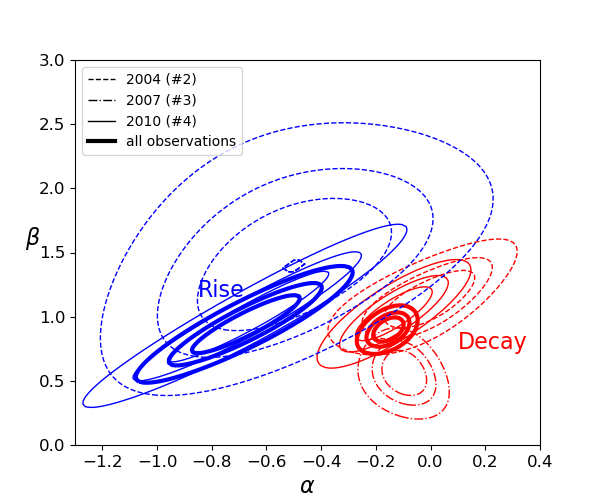}}
    \caption{ Contour plots $\beta$-$\alpha$ for the rising (blue) and decaying (red) phases of the outburst of 2004 (\#2, thin dashed line), 2007 (\#3, thin dot-dashed line) and 2010 (\#4, thin solid line). The  contours in thick solid lines represent the dependency when fitting all rising (blue) or decaying (red) phase radio fluxes simultaneously. Confidence contour levels correspond to 68\%, 90\% and 99\% ($\Delta\chi^2$ of $2.3$, $4.61$ and $9.2$ respectively). The contours are obtained when fitting only the quasi-simultaneous radio/X-ray observations (not the interpolated radio observations).}
    \label{fig:contours}
\end{figure}

We first test Eq. (\ref{eq:Fradio1}) in the rising phase of outburst \#4. 
We fit all 16 radio observations with simultaneous or quasi-simultaneous radio/X-ray data in {\sc{xspec}}. We set $r_J$ and \mdot of each observation to the best quasi-simultaneous X-ray fit values\footnote{When fitting simultaneously X-ray and radio data, if the JED-SAD and reflections parameters are left free to vary simultaneously to $\Tilde{f}^{*}$, $\alpha$ and $\beta$, the X-ray fit is found to be significantly worse, especially around the iron line, for the benefit of a perfect match of the radio fluxes. By freezing the JED-SAD and reflections parameters to their best fit values obtained by fitting the X-rays, we rather chose to favor the X-ray fit for which we have a fully developed physically motivated spectral model.} (obtained in Sect. \ref{sec:norad}). We implement in {\sc{xspec}} a model to fit the radio emission following Eq. (\ref{eq:Fradio1}). We impose the same value of $\Tilde{f}^{*}$, $\alpha$ and $\beta$ between all the observations. The number of radio fluxes we use can be found in Table \ref{tab:mjd_hard}.

As the errors on the radio are sometimes quite small, we introduce a 10\% systematic errors\footnote{This is done so that the fit is not driven by one radio flux only but tries to reproduce all the fluxes within this 10\% error margin. We note that the maximum variation observed in the radio light-curve is about 20\% variation within three days (see Fig \ref{fig:Radio_interp}). Thus within the one day delay between the radio and X-ray observations, we do not expect variations exceeding the 10\% systematic error we add, justifying the use of non-exactly simultaneous X-ray and radio pointings. The effects of adding systematic errors is discussed in appendix \ref{sec:c_systematics}.} to the radio fluxes to account for the non simultaneity between the radio and X-ray observations and few percent radio intrinsic variability (\citealt{cor00}).
 
\begin{figure}[!]
	\includegraphics[width=\columnwidth]{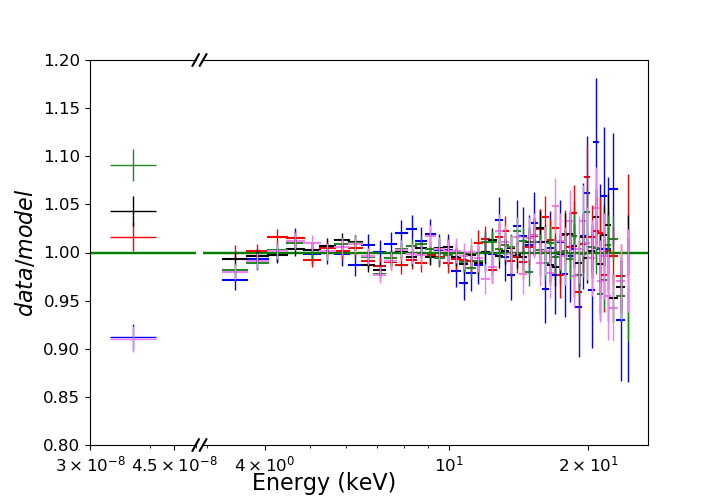}
    \caption{Ratio between the data and the model for the best fit of 5 among the 16 multiwavelength observations (radio/X-ray) of the rising phase of the 2010 outburst (MJD 55217 in blue, 55259 in red, 55271 in black, 55288 in green and 55292 in violet). We choose to show only 5 ratios to ease the plot but the best fit was obtained by using all the simultaneous or quasi-simultaneous radio/X-ray observation, fixing the JED-SAD parameters to the best fitting values obtained by fitting the X-ray spectra first, and then fitting the radio points with Eq. (\ref{eq:Fradio1}).}
    \label{Fig:rise_residuals}
\end{figure}

The best fit gives 
$\Tilde{f}^{*}= 7.1^{+\,15}_{-5}\times10^{-8}$, $\alpha= -0.66 \pm 0.32 $ and $\beta= 1.00^{+0.39}_{-0.38}$.
The contours $\alpha$-$\beta$ are also reported as blue thin solid lines in Fig. \ref{fig:contours}. The fit reproduces all the radio fluxes within an error lower than 10\% (see examples of residuals in Fig. \ref{Fig:rise_residuals}) suggesting that Eq. (\ref{eq:Fradio1}) works adequately. The positive value of $\beta$ is consistent with the observed correlation between the radio emission and the luminosity of the binary system. Concerning the negative value of $\alpha$, it agrees with a decrease of the inner radius of the SAD when the system reaches bright hard states with stronger radio emission as expected in our JED-SAD approach (and similarly to most of the truncated disk models like \citealt{esin1997advection}).

\begin{figure*}[t]
   \centering
            { \includegraphics[width=\textwidth]{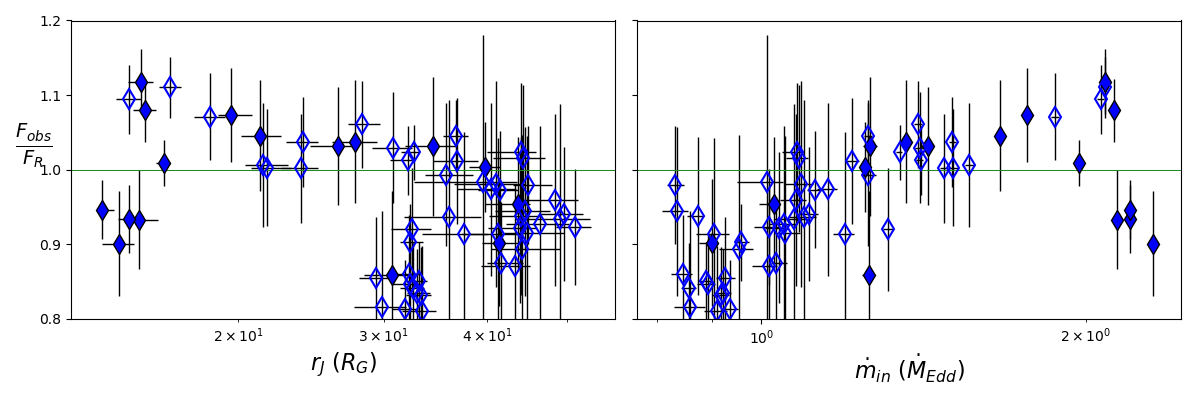}}
            
      \caption{Ratio between all radio fluxes of the rising phase of outburst \#4 and the results of Eq. (\ref{eq:Fradio1}) as function of \rJ (left) and \mdot (right). The filled blue points represent the 16 quasi-simultaneous radio/X-ray observations, while the empty points represent the interpolated radio fluxes (see Appendix \ref{sec:a_interp}). The radio observations are well reproduced using the values $\Tilde{f}^{*}=7.1\times10^{-8}$, $\alpha= -0.66$ and $\beta= 1.00$. }
         \label{fig:ftilde_free}
   \end{figure*}

In a second step, we apply the same procedure to all the interpolated radio fluxes of the rising phase of outburst \#4. Following the first step,
we set \rJ and \mdot to their best fit values obtained when fitting the X-ray alone. Then we reproduce the radio using the best fit values of $\Tilde{f}^{*}$, $\alpha$ and $\beta$ obtained previously to compute the expected radio flux $F_R$ using Eq. (\ref{eq:Fradio1}). The corresponding ratios $F_{obs}/F_R$ are reported in Fig. \ref{fig:ftilde_free}. There is almost no remaining dependency on \rJ or $\dot{m}_{in}$. Compared to Fig. \ref{fig:ftilderj}, this now shows a much clustered distribution around 1, with a dispersion of about $\pm15\%$. 

\begin{figure}[!]
	\includegraphics[width=\columnwidth]{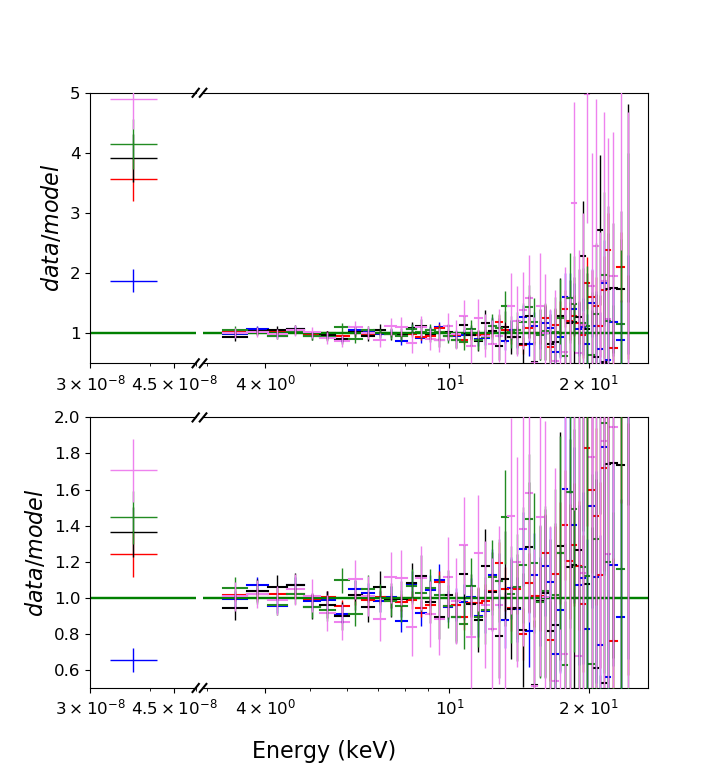}
    \caption{Ratio between the data and the model for the best fit of five among the eight multiwavelength observations of the decaying phase of outburst \#4 (MJD 55613 in blue, 55617 in red, 55620 in black, 55630 in green and 55639 in violet). \textbf{Top panel}: All fits were done simultaneously, fixing the parameters of Eq. (\ref{eq:Fradio1}) to the ones found in the rising phase : $\Tilde{f}^{*}= 7.1\times10^{-8}$, $\alpha= -0.66$ and $\beta= 1.00$. 
    \textbf{Bottom panel}: fixing $\alpha$ and $\beta$ to the values found for the rising phase. $\Tilde{f}^*$ is free to vary and converges to the value $2.0\times10^{-7}$.} %
    \label{fig:ratiodecay}
\end{figure}

\subsection{Decaying phase of the 2010 outburst}
For the decaying phase of outburst \#4, we proceed similarly to the rising phase. We chose all eight observations, with simultaneous or quasi (less than 1 day) simultaneous radio/X-ray observations of the decaying phase of the outburst \#4. We set the values of $r_J$ and \mdot to the best X-ray fit, then we fit the radio fluxes using Eq. (\ref{eq:Fradio1}).

As a first test, we set $\Tilde{f}^{*}$, $\alpha$ and $\beta$ to the best fit values obtained in the rising phase. The corresponding data/model ratio is plotted in Fig. \ref{fig:ratiodecay}. The top panel shows that using the value of $\tilde{f}^*$ of the rising phase in the decaying phase induces an error in the radio flux up to a factor of 5.
The bottom panel shows that even if we let the scaling factor $\tilde{f}^*$ free, converging to the value $2.0\times10^{-7}$, the radio flux is wrong by a factor up to 1.8.
Thus, the parameters $\Tilde{f}^{*}$, $\alpha$ and $\beta$ cannot be the same as the ones obtained in the rising phase.\\

Following what we present for the rising phase, we now let $\Tilde{f}^{*}$, $\alpha$ and $\beta$ free to vary but tied between all the observations. The best fit values are 
$\Tilde{f}^{*}= 2.9_{-0.9}^{+1.0}\times10^{-8}$, $\alpha= -0.13_{-0.16}^{+0.15}$ and $\beta= 1.02\pm0.23$. The corresponding confidence contour $\alpha$-$\beta$ is plotted as red thin solid lines in Fig. \ref{fig:contours}. It is clearly inconsistent with the blue contour obtained in the rising phase.

\subsection{Comparison with the other outbursts}
We constrain the functional dependency of the radio emission of the other outbursts by repeating a similar analysis. We thus need at least three observations taken in the corresponding rising and decaying phases to constrain the three free parameters $\alpha$, $\beta$ and $\Tilde{f}^{*}$. Only outburst \#2 (year 2004) and the decaying phase of \#3 (year 2007) have the sufficient number of simultaneous/quasi simultaneous radio and X-ray observations to apply our procedure. The number of radio fluxes we use for each phase of the outbursts can be found in Table \ref{tab:mjd_hard}. The corresponding contour plots of $\alpha$-$\beta$ are overplotted in Fig. \ref{fig:contours} in dashed and dot-dashed lines respectively. Two results are remarkable. First, and similarly to outburst \#4, we need different functional dependencies of the radio emission with $r_J$ and $\dot{m}_{in}$ between the rising and decaying phase for outburst \#2. Even more interestingly, the values obtained for $\alpha$ and $\beta$ are in quite good agreement between the different outbursts, the contour of the decaying phase of outburst \#3 also  close to the contours of  the decaying phases of outbursts \#2 and \#4. 

While this could be rather surprising given the quite simple expression used to model the radio emission, we believe that this result reveals intrinsic differences in the jet emission origin (see Sect. \ref{Sec:discuss} for this discussion).\\

In a last step, we use all the quasi-simultaneous radio observations, simultaneously fitting all the rising phase observations together with the same parameters $\alpha$ and $\beta$ for all outbursts but with different normalization $\Tilde{f}^{*}$ for each outburst. We did the same for all the decaying phase observations. The resulting $\alpha$-$\beta$ contours have been plotted in Fig. \ref{fig:contours} in thick solid lines. 

It confirms the two different, and mutually inconsistent, functional dependencies of the radio emission on \rJ and \mdot between the rising and decaying phases observations. While the radio flux observed in the rising phases is well reproduced (within about 15\%) by the relation 
\begin{equation}
    F_{R}^{rise} \propto r_J^{{-0.67}_{-0.22}^{+0.21}}\, \dot{m}_{in}^{{0.94}_{-0.24}^{+0.25}} 
    \label{eq:Fradio1rise}    
\end{equation}
in decaying phases it rather follows
\begin{equation}
    F_{R}^{decay} \propto r_J^{-0.15\pm 0.06}\, \, \dot{m}_{in}^{{0.9}\pm0.1} 
    \label{eq:Fradio1decay}    
\end{equation}
with a weaker dependency on $r_J$.\\

\begin{figure*}[!]
    {\includegraphics[width=\textwidth]{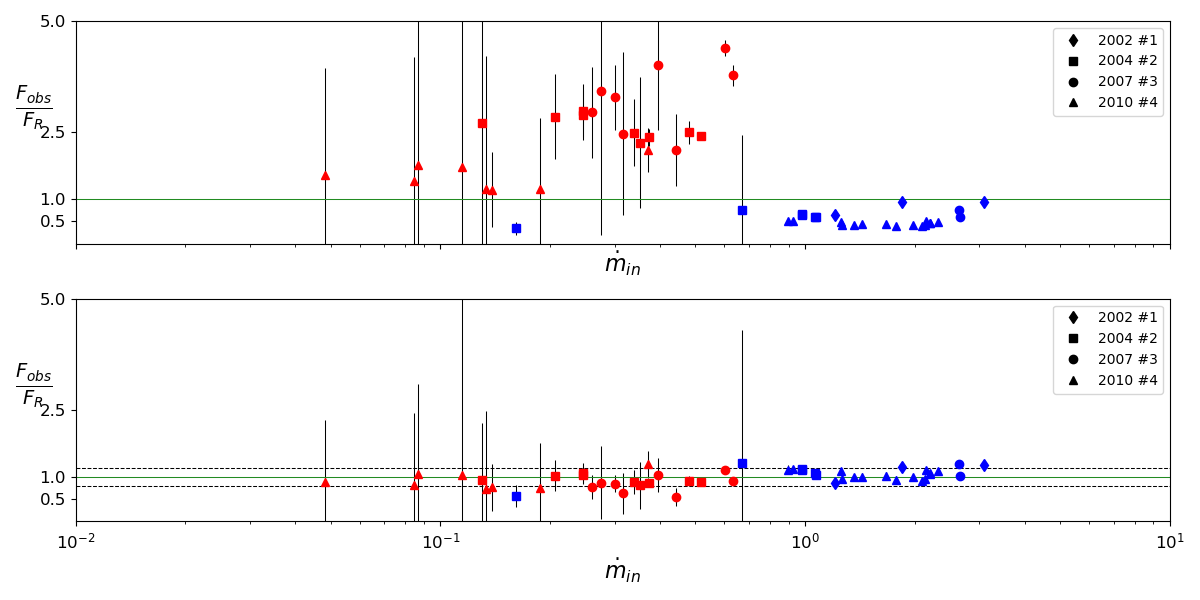}}
    \caption{Ratios between the observed radio fluxes and the modeled radio fluxes for all the quasi-simultaneous radio/X-ray observations of the 4 outbursts. Each outburst is represented using a different marker. In blue, the rising phases and in red the decaying phases. The modeled radio fluxes have been obtained using Eq. (\ref{eq:Fradio1}). The parameters are  ($\alpha=-0.67$, $\beta=0.94$) for the rising phases and ($\alpha=-0.15$, $\beta=0.9$)\ for the decaying phases. \textbf{Top panel}: We use  $\tilde{f}^*=4.1\times10^{-8}$ for all outbursts. \textbf{Bottom panel}: We use different $\tilde{f}^*$ for each phase of the outbursts. All values of $\tilde{f}^*$ used are reported in Table \ref{tab:ftilde}. The horizontal dashed lines represent a 20\% error margin (ratio of 0.8 and 1.2 respectively).  }
    \label{fig:ratios_radio}
\end{figure*}

Some variations of $\Tilde{f}^{*}$ are however required to significantly improve the radio emission modeling. This can be seen in Fig. \ref{fig:ratios_radio}, where we report the ratios $F_{obs}/F_R$ using Eq. (\ref{eq:Fradio1rise}) to compute the radio flux if the observation is in the rising phase and Eq. (\ref{eq:Fradio1decay}) if in the decaying phase. At the top we use the same value $\tilde{f}^*$ for all outbursts. While the ratios cluster around 1 there is some scattering between the different phases of the different outbursts. We report in the bottom panel of Fig. \ref{fig:ratios_radio} the same ratio but letting  $\tilde{f}^*$ free to vary between outbursts and between the rising and decaying phases. The improvement is clear and almost all radio fluxes can be reproduced within a 20 \% margin error. The different values of $\tilde{f}^*$ found are reported in Table \ref{tab:ftilde}. We observe variation up to a factor three (e.g., between the rising phase of the 2002 and 2010 outbursts). This could be related to local changes of the radiative efficiency of the radio emission from outburst to outburst.

   \begin{table}[h!]
      \caption[]{Values of $\tilde{f}^*$ found for each phase of the outbursts. Obtained when fitting all the quasi-simultaneous observations simultaneously.}
         \label{tab:ftilde}
         \centering
         \begin{tabular}{|c|c|c|}
            \hline
            Outburst & Rise & Decay \\
            \hline
            2002 \#1 & $4.1^{+5.4}_{-2.3}\times10^{-8}$ & -  \\ \hline
            2004 \#2 & $5.5^{+6.6}_{-2.9}\times10^{-8}$ & $1.4^{+0.3}_{-0.3}\times10^{-8}$ \\ \hline
            2007 \#3 & $5.3^{+6.6}_{-2.9}\times10^{-8}$ & $1.0^{+0.3}_{-0.2}\times10^{-8}$\\ \hline
            2010 \#4 & $7.2^{+9.2}_{-3.7}\times10^{-8}$ & $2.3^{+0.7}_{-0.5}\times10^{-8}$ \\
            \hline
         \end{tabular}
   \end{table}

\vspace*{-0.2cm}

\section{Discussion} \label{Sec:discuss}

We present in this paper the first X-ray spectral fits of an X-ray binary using the JED-SAD model. Compared to previous works, we constructed model tables that enable the use of {\sc{xspec}} for a direct fit procedure. We also constructed a reflection table, based on {\sc{relxill}}, using as inputs the photon index and high-energy cutoff that best fit the JED-SAD spectral shapes. We obtain good fits for all the X-ray observations of GX-339-4 during the hard states observed by RXTE on the period 2002-2010 by only varying the accretion rate $\dot{m}_{in}$ of our system and the transition radius $r_J$ between the JED and the SAD. 

As said in the introduction, in the absence of a known physical law that would link these two parameters, we let them free to vary independently one with each other in the fit procedure. This is the simplest approach to try to understand, and hopefully to physically interpret (see Ferreira et al. in preparation), their behavior.

Then, radio emissions simultaneous (or quasi simultaneous by one day) to the X-rays were reproduced using a generic formula only depending on $\dot{m}_{in}$ and $r_J$. One of the main results of this spectral analysis is the necessity of a different functional dependency of the radio emission on $r_J$ and $\dot{m}_{in}$. The two different expressions for the radio emission are reported in Eqs. \ref{eq:Fradio1rise} and (\ref{eq:Fradio1decay}). We believe that this difference in functional dependency is a "back product" of the true physical link between \rJ and $\dot{m}_{in}$.

\subsection{Indications of different radiative behaviors between the rising and decaying phases}

Observational clues on the jet behavior can be derived from a set of different diagnostics: (i) the radio spectral index $\alpha_R$, (ii) the measure of the spectral break frequency $\nu_{break}$, (iii) timing properties, (iv) the correlation $L_R-L_X$ and (v) linking the radio luminosity $L_R$ to disk properties ($\dot{m}_{in}, r_J$). Items (i)-(iv) are discussed in this section while the item (V) is discussed in Sect. \ref{sectBPvsBZ}.

The radio spectral index  $\alpha_R$ can be analytically derived under the assumption of a self-absorbed synchrotron emission smoothly distributed along the jet. It depends on the particle distribution function, the jet geometry and the way the dominant magnetic field varies with the distance (see Eq A.8 in Appendix of \citealt{marcel2018aunified}). There is a priori no reason to assume that these parameters should not vary in time.
Observationally, there is however no clear evidence of differences in the radio spectral index $\alpha_R$ between the rising and decaying phases of GX 339-4 (Espinasse private communication, see also \citealt{koljonen2019} for more detailed discussion on this point and \citealt{tremou20} for the quiescent state case where the radio spectrum is clearly inverted). Although this is already an important information, we note nevertheless that these $\alpha_R$ are derived within a rather limited  radio band and might therefore not be fully representative of the whole jet spectrum (see for instance \citealt{peault2019modelling}).

\begin{figure}[t]
	\includegraphics[width=\columnwidth]{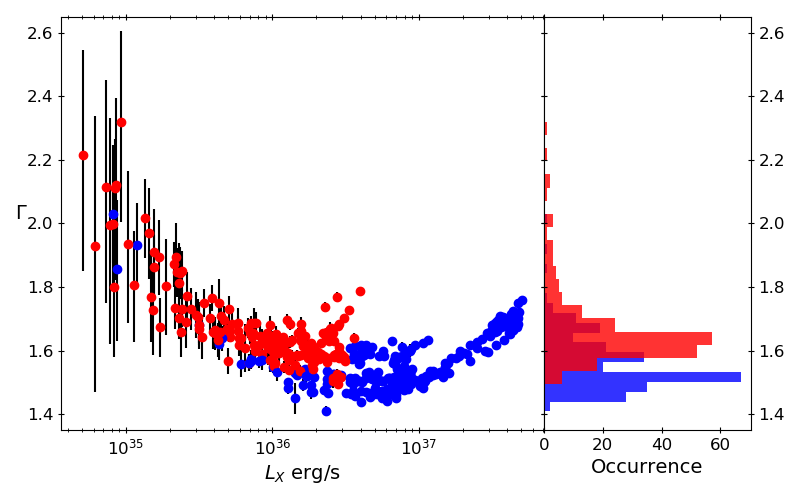}
    \caption{\textbf{Left:} Hard X-ray power law index $\Gamma$ as a function of the 3-9 keV X-ray luminosity of the pure hard state observations during the four GX339-4 outbursts (data from \citealt{clavel2016systematic}). In blue the rising phase and in red the decaying phase. \textbf{Right:} Histograms of $\Gamma$. These distributions are however subject to a certain number of observational biases: inclusion of error bars and the number of observations done per phase.}
    \label{fig:gamma}
\end{figure}

The evolution of the spectral break frequency $\nu_{break}$, marking the transition from self-absorbed to optically-thin jet synchrotron radiation, could however be different in the two (rising and decaying) phases. The radio spectral index being  flat or inverted in the hard state, the power of the jets is mainly sensitive to the position
of the spectral break. \cite{gandhi2011variable} measured this break at $\sim$5$\times 10^{13}$ Hz in a bright hard state during the rise of the 2010-2011 outburst. By comparison, \cite{cor13} constrain the break to be at lower frequency in the decaying phase,  suggesting a less powerful jet in this phase.
There is also a potential link between the X-ray hardness and the jet spectral break frequency, harder X-ray spectra having a higher $\nu_{break}$ (\citealt{russell2014accretion,koljonen2015connection}). Interestingly, GX 339-4 shows on average a softer power-law index in the decaying phase compared to the rising phase (see Fig. 
\ref{fig:gamma}). Given the observed correlation between $\nu_{break}$ and the X-ray hardness, this also suggests a different behavior for $\nu_{break}$ (and consequently of the jet power) 
between the two phases.

There are other indications that the accretion (through the X-rays emission) and ejection (through the radio emission) processes could behave differently at the beginning and the end of the outburst. At first sight, the radio/X-ray correlation followed by GX339-4 agrees with a linear correlation of index $\sim$0.7 in log-log space (e.g., \citealt{cor00,cor03,cor13}) even down to very quiescent states \citep{tremou20}. But a more careful analysis shows the presence of wiggles along this linear correlation especially between the high and low luminosity states (e.g., \citealt{cor13}, Fig. 8). When looking more precisely to the rising and decaying phase, two different correlations may even be observed \citep{islam18}.

These differences may be linked to a change of the radiative efficiency of the X-ray corona with luminosity. Indeed, the low X-ray luminosity states, below 2-20\% of the Eddington luminosity, are potentially less radiatively-efficient than the high X-ray luminosity states (\citealt{koljonen2019}; Marcel et al. in preparation). As noticed by \cite{koljonen2019}, these changes of the accretion flow properties could affect the jet launching and therefore its radio emission properties.

\begin{figure}[t]
	\includegraphics[width=\columnwidth]{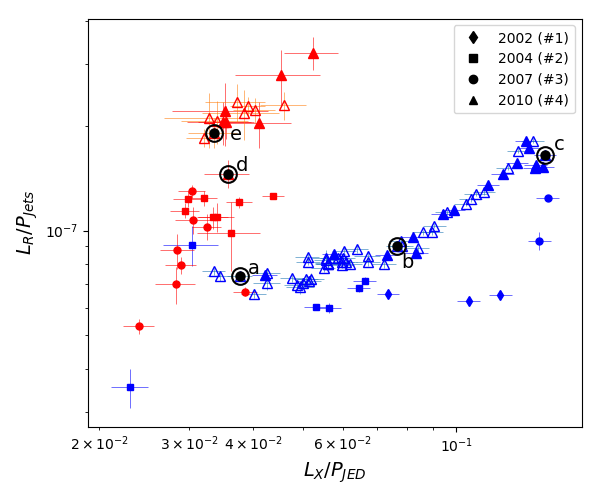}
    \caption{Radio emission efficiency ($L_{R}=L_{9 GHz}/P_{Jets}$) versus X-ray emission efficiency ($L_{X}=L_{3-9 keV}/P_{JED}$)  during the outbursts of GX339-4. The blue points are the rising phases. The red points are the decaying phases. The markers serves to distinguish the different outbursts: diamonds for 2002, squares for 2004, dots for 2007 and triangle for 2010 (filled for the quasi-simultaneous observations and empty for the interpolated radio fluxes). We highlighted the 5 observations (marked \textit{a} to \textit{e}) presented in Fig. \ref{Fig:showing_fits} to provide the chronological evolution of an outburst.}
    \label{fig:puissance}
\end{figure}

In the JED-SAD model, the accretion power available in the accretion flow, $\displaystyle P_{acc}=\frac{GM\dot{M}}{2R_{isco}}\left [1-\left (\frac{r_{isco}}{r_J}\right )^{1-p}\right ]$ (see Sect. \ref{introJEDSAD} for the definition of $p$ and $b$), is released in three different form: advection, radiation and ejection. The first two happening inside the JED and their sum is defined as $P_{JED}=(1-b)\, P_{acc}$. The ejection power is released in the jets and is defined as $P_{jets}= b\, P_{acc}$. We also define the ratio $\eta_{R}=L_{R}/P_{jets}$ and the ratio $\eta_X=L_{3-9\,keV}/P_{JED}$ that can be respectively interpreted as the radiative efficiency in the radio and X-ray bands. We report in Fig. \ref{fig:puissance} the ratio $\eta_{R}$\footnote{In the case of the 2010 outburst, the full triangle are quasi-simultaneous radio fluxes whereas the empty triangle use the interpolated radio luminosity $L_{R}$ computed for all the X-ray observations (see Fig. \ref{fig:Radio_interp}).} as function of the ratio $\eta_X$ for the rising (blue points) and the decaying phase (red points) of the outbursts.  We highlight in Fig. \ref{fig:puissance} the observations (labeled a to e) presented in Fig. \ref{Fig:showing_fits} to mark the chronological evolution along an outburst. Figure \ref{fig:puissance} mostly depends on the well constrained mass accretion rate obtained with our fits of each X-ray observations.

The blue points of the rising phases follow a similar trend for all the outbursts with a change of the X-ray and radio radiative efficiency by a factor $\sim$4 and $\sim$2 respectively.
In the decaying phase however, 
each outburst clusters at a same radio and X-ray radiative efficiency. Interestingly, the radio radiative efficiency changes from outburst to outburst while the X-ray radiative efficiency stay roughly constant at $\eta_X\sim 3-4 \times 10^{-2}$, the lowest values observed in the rising phase. These results suggest indeed a change of the radiative properties of the accretion-ejection structure between the beginning and the end of the outburst. And it is possible that it has some impact on the functional dependency of the radio emission highlighted in this paper. Contrary to the conclusion of \cite{koljonen2019} however, the accretion rate does not seem to be the (unique?) parameter that controls the evolution of $\eta_R$. Indeed, looking at outburst \#2 and \#4 separately, $\eta_R$ stays roughly constant in the decaying phase of each outburst, whereas $\dot{m}_{in}$ varies by at least a factor 10 (see Fig. \ref{Fig:Fits_results_norad}). And different radio efficiencies are observed between each outburst during the decaying phases even at similar values of $\dot{m}_{in}$. Something else seems to be also at work.

\subsection{Changes of the dynamical ejection properties?}
\label{sectBPvsBZ}

The existence of two functional dependencies $F_R(\dot{m}_{in}, r_J)$ raises a profound question. Radiative processes in jets are local and are independent of disk parameters such as $\dot{m}_{in}$ and $r_J$. However the fact that the time evolution $F_R(t)$ can be quite accurately reproduced with a function of ($\dot{m}_{in}, r_J$) shows that global jet parameters do actually depend on them. These parameters, which constitute the jet dynamics, are for instance the magnetic field strength and geometry, the jet collimation degree, the existence of internal chocs or even jet instabilities. Our findings seem therefore to highlight two different jet dynamics.

\subsubsection{A threshold in $\dot{m}_{in}$?}

The hard state data sets of the rising and decaying phases used in this analysis do not overlap in terms of accretion rate. Only two out of the 28 radio observations of the rising phases require a mass accretion rate comparable to those observed during the decaying phases.  All the others have a higher mass accretion rate than the decaying phases. This is of course an observational bias due to the difficulties to catch the source as fast as possible at the beginning of the outburst.
But the detected difference in the functional dependency of $F_R$ could be due to some threshold in $\dot{m}_{in}$ that could, in turn, translate into some difference in the way the radio emission scales with the disk parameters. Above the threshold, the radio emission would follow Eq. (\ref{eq:Fradio1rise}) and below, Eq. (\ref{eq:Fradio1decay}).  Given the too small number of low accretion rate hard states in the rising phases,  our analysis cannot test this possibility. Clearly, more observations are needed to assess this hypothesis. 

However we do not favor this interpretation. The main reason is that the rising and decaying hard states are temporally disconnected. The source stays several months in the soft state between these two hard state phases. Thus they do not share the same "history". The hard states in the rising phase come from a quiescent, already radio emitting, state while the hard states in the decaying phase come from soft, radio silent, states. This rather supports a link with the global jet structure (as proposed in the Sect. \ref{sec:ejection_process}) rather than a threshold in $\dot{m}_{in}$.

\subsubsection{A change in the dominating ejection process?}
\label{sec:ejection_process}
Since the commonly invoked radiative process is synchrotron, the first thing that comes to mind to explain this difference is the magnetic field strength. The only reasonable assumption to make is that this field is proportional to the field anchored at the JED, which writes (see \citealt{marcel2018aunified} for more details):
\begin{equation}
B_z(r)=(\mu\,\mu_0\, P_{tot})^{1/2} \simeq \Bigg(\mu \,\mu_0\, P_* \frac{\dot{m}_{in} r^{-5/2}}{m_s}\Bigg)^{1/2}
\label{eqB}
\end{equation}
where $\mu$ is the magnetization, $m_s$ the accretion Mach number, $P_* = m_i n_* c^2$ and $n_* = \frac{1}{\sigma_T R_g}$. Assuming constant JED parameters $\mu=0.5$ and $m_s=1.5$ used in our model, we evaluate the magnetic field strength measured in $r_{isco}$ at around $10^8$ G during the outbursts.

Note that a JED exists within a small interval $[\mu_{min}, \mu_{max}]$ of disk magnetization $\mu$, with $\mu_{min}\sim 0.1$ and $\mu_{max}\sim 0.8$ \citep{fer97}.  
The existence of such an interval has led \citet{petrucci2008role} to propose that the hysteresis observed in XrBs could be a consequence of a JED switch-off with $\mu=\mu_{min}$ and switch-on at $\mu=\mu_{max}$. 
In the spectral analysis shown in the present paper, we have supposed a constant $\mu$ since, as shown in \citet{marcel2018aunified}, the JED spectra are poorly affected by $\mu$ within the allowed parameter space. However, the possible difference in magnetisation between the rising and decaying phase could also have a direct impact on the jet dynamical and radiative properties, explaining the change of the observed radio behavior. According to Eq. (\ref{eqB}), a dichotomy of the magnetization $\mu$ at a given value of the mass accretion rate $\dot{m}_{in}$ entails a dichotomy in the magnetic field strength. Thus the rising phase, switching-off with $\mu=\mu_{min}$, would present a weaker magnetic field strength compared to the decaying phase, switching-on with $\mu=\mu_{max}$. This difference in the magnetic field strength could play a role in the difference of functional dependency of the radio emission. This could also explain the higher radio efficiencies observed in Fig. \ref{fig:puissance} during the decaying phases (e.g., \citealt{cas09}). \\

Another possibility could be suggested by the most recent numerical simulations showing that the vertical magnetic field is carried in and accumulates around the black hole (building up a magnetic flux  $\Phi_{bh}$) until the surrounding disk magnetization reaches a maximal value near unity (see e.g., \citealt{tchek2011, Liska2020}). In our view, the inner disk regions are nothing else than a Jet Emitting Disk driving a Blandford \& Payne jet (BP hereafter, \citealt{blandford1982hydromagnetic}), although a Blandford \& Znajek spine (BZ hereafter, \citealt{blandford1977electromagnetic}) launched at its midst has attracted more attention in the literature\footnote{The inner disk regions have been usually termed MAD for Magnetically Arrested accretion Disk \citep{narayan2003magnetically,tchek2011}. But as accurately noticed by \citet{mckinney2012general}, a thin or even slim disk is not arrested. The deviation from a Keplerian rotation is only of the order the disk thickness and its structure resembles the JED, with a near equipartition magnetic field.}. 
Then another possible explanation for the existence of two functional dependencies for $F_R(\dot{m}_{in}, r_J)$ could be that jets are two-component MHD outflows: a BZ spine, taping the rotational energy of the black hole, surrounded by a BP jet, taping the accretion energy reservoir of the disk. The jet dynamics and subsequent radio emission then depend on the relative importance of these two flows, that can be roughly measured by the ratio of the magnetic flux associated to each component, namely $\Phi_{bh}$ for the spine and $\Phi_{JED}$ for the outer BP jet. By construction, $\Phi_{bh}$ builds upon $\Phi_{JED}$ and reaches large values, such as $\Tilde{\Phi}_{bh}=\Phi_{bh}/(< \dot{M}_{in} > r_g^2c)^{1/2}\sim 50$, only if a large magnetic flux is available initially in the disk \citep{tchek2011, Liska2020}. The functional dependency with $r_J$ that we observe for the radio flux in our fits of the rising phase spectra could thus come from the dependency of $\Tilde{\Phi}_{bh}$ on $r_J$.\\

\begin{figure}[t]
	\includegraphics[width=\columnwidth]{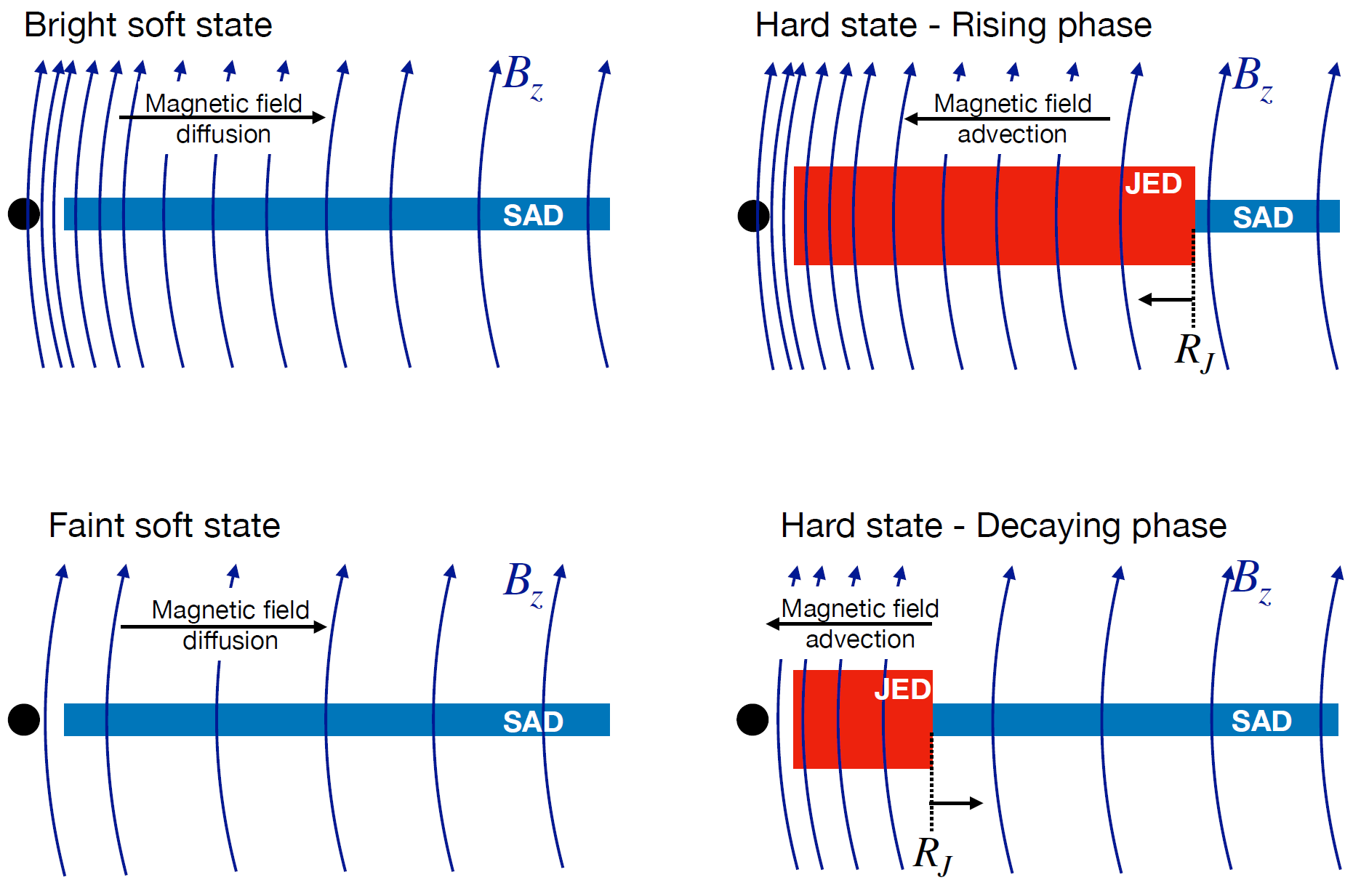}
    \caption{Sketches of the inner regions of an accretion flow around a black hole during the different phases of the outburst. {\bf Top right:} In the rising hard phase, the JED is settled over a large region ($R_J$ is large), leading to efficient magnetic flux accumulation on the black hole. The radio emission arises from a two-component outflow, made of an important BZ spine surrounded by a BP jet.  
{\bf Top left to bottom left:} During the soft state, there is no more JED, the magnetic field diffuses away and both BP and BZ jets disappear.
No or weak radio emission is expected (jetline). {\bf Bottom right:} In the decaying phase, a JED reappears in the innermost region, and the magnetic field advection becomes efficient again. The magnetic flux on the black hole is still weak however and the BZ spine has little or no impact on the jet dynamics and subsequent radio emission. 
}
    \label{fig:BZBP}
\end{figure}

A simple scenario can then be designed and is sketched in Fig. \ref{fig:BZBP}. During the rising hard state phase, $r_J$ is initially large and decreases in time (top right). The system comes from a quiescent state and the presence of a JED over a large radial extent allowed the disk to build up a maximal  $\Phi_{bh}$. The spine is very important and affects the overall jet dynamics, which translates into a radio flux described by Eq. (\ref{eq:Fradio1rise}). When the disk magnetization becomes too small, the JED transits to a SAD accretion mode (left, top and bottom). The magnetic field diffuses away, decreasing thereby $\Phi_{bh}$ and no more jets are observed (neither BP nor BZ). As long as the system remains in the soft state, the field keeps on diffusing away until some equilibrium is eventually reached. At some point however, the outburst declines which translates into a decrease of the inner disk pressure and, thereby, an increase of the disk magnetization. In this decaying phase, an inner JED becomes re-ignited inside-out, with its bipolar BP jets but with a limited magnetic flux available (bottom right). By construction, $\Phi_{bh}$ remains small and the BZ spine has a limited impact on the overall jet dynamics. That would translate into a radio flux described by Eq. (\ref{eq:Fradio1decay}), until the JED is rebuilt over a large enough radial extent.\\

There are many uncertainties in our different interpretations, since our JED-SAD modeling has its own simplifications.
This last scenario is only a tentative to provide an explanation to our puzzling finding. Quite interestingly it provides also a means to observationally test it. Indeed, it relies on the existence of a BZ spine in the case of GX 339-4, which is a black hole candidate. Around a neutron star, the invoked scenario of magnetic flux accumulation into the central object should obviously not work. It would therefore be useful to investigate any changes in the radio properties during the rise and decay phases for neutron star binaries.

This scenario may look similar to the ones proposed by \cite{beg14} or \cite{kyl15} where the presence of a hot inner corona (an ADAF-like accretion flow in both cases) would help in accumulating/creating the required magnetic field that will eventually produce a jet. However in these two approaches it is not clear why the process would differ between the rising and decaying phases and how the functional dependency of the radio emission would depend on the ADAF properties. Clearly, more dedicated works should be done in this respect.\\

\subsection{Effects of the JED-SAD parameters}

In the results shown in this work, we use the same values of the JED-SAD parameters $b$, $m_s$ and $p$ as in M19 and M20. A detailed study of the JED-SAD parameter space has already been performed (see \citealt{marcel2018aunified} Sect. 4) and converges on these values in the case of GX 339-4. The parameter $p$ has almost no spectral impact in the \textit{RXTE}/PCA energy range (3-25 keV), this can be seen in Fig. 10 of \cite{marcel2018aunified} ($p$ was called $\xi$ at the time). Thus, we do not expect any variation in the fits. Letting p free would result in an unconstrained parameter. The main impact of the parameters $b$ and $m_s$ is a variation of the maximum temperature in the JED and results in a variation of the high-energy cutoff of the hard X-ray emission. However, the high-energy cutoff is not visible within the \textit{RXTE}/PCA energy range used in our fitting procedure. We are then unable to constrain these parameters from the data and choose to set them to the same values used by M19 and M20. This also allows us to compare the evolution of the main JED-SAD parameters $r_J$ and $\dot{m}_{in}$ with the qualitative results obtained by M19 and M20.

However, we study the impact of using other values of $m_s$ or $b$. The global trend of the parameters stays similar, with only slightly different values of $r_J$ and $\dot{m}_{in}$ (see Fig. \ref{fig:ms_2010} and \ref{fig:b_2010} in appendix). The biggest difference are observed with the different values of $m_s$. We thus concentrate on this parameter to see the effect of its value on the $\alpha-\beta$ contours. They are reported in Fig. \ref{fig:contours_diff_ms}. The contour plots of the rising and decaying phase vary for different values of $m_s$ however, they are never consistent between each other making our main conclusions unchanged: we need two different functional dependencies between the rising and decaying phases.

\section{Conclusion} \label{sec:conclusion}

We have presented in this paper the first direct fit of the X-ray data of an X-ray binary with our JED-SAD model. We have constructed fits format tables that can be used in {\sc{xspec}}. This includes a reflection table based on the {\sc{xillver}} reflection model (\citealt{garcia2013x}). We applied our model to the X-ray observations of GX339-4, focusing on the "pure" hard-states phases of the 4 outbursts observed during the \textit{RXTE} lifetime. We have deduced from the fits the temporal evolution of the main parameters of the JED-SAD configurations that is the inner accretion rate $\dot{m}_{in}$ and the transition radius $r_J$ between the inner JED and the outer SAD (see Fig. \ref{Fig:Fits_results_norad}). This evolution is in relatively good agreement with the qualitative estimates done by M19 and M20, our spectral fit procedure putting however much stronger constraints especially on $r_J$.\\

We were also able to put constraints on the functional dependency of the radio emission with $r_J$ and $\dot{m}_{in}$ for all outbursts of GX 339-4. Assuming a general radio flux expression $F_R= \tilde{f} \dot{m}_{in}^{\beta} r_J^{\alpha}$, we were able to constrain the values of $\alpha$ and $\beta$ for the different outbursts. These values appear consistent between the different rising phases or the different decaying phases. But the rising/high mass accretion rate and decaying/low accretion rate phases solutions differ significantly. In the rising phase, the radio emission varies as $\sim\dot{m}_{in}^{1} r_J^{-0.6}$ while in the decaying phase the radio emission has a weaker dependency with $r_J$ and varies as $\sim\dot{m}_{in}^{0.9}r_J^{-0.2}$. While the exact values of the indexes depends slightly on the JED-SAD parameters, the two solutions obtained for the rising and the decaying phases are always mutually exclusive. A significant improvement of the fit of the radio fluxes is obtained by letting the scaling factor $\tilde{f}$ free to vary between the different phases. The observed variation (up to a factor 2) could correspond to a change of the radiative efficiency of the radio emitting process from outburst to outburst. \\

We suggest a few explanations for the difference in the functional dependency of the radio emission with $r_J$ and $\dot{m}_{in}$ between the rise and decay phases. A clear understanding appear challenging given the scarce information we have on crucial jet parameters like the magnetic field strength and geometry, the jet collimation degree, the existence of internal chocs or even jet instabilities. A possible scenario relies on a change of the relative importance of the Blandford-Znajeck vs Blandford \& Payne processes in the radio emitting process due to the expected evolution of the  magnetic field strength in the inner part of the accretion flow.\\

All results of this paper should be tested on other XrB and on more recent outbursts of GX339-4. It would be also interesting to see how the case of neutron stars (where no BZ is expected) or objects belonging to the so-called \textit{outlier} population (\citealt{cori11}) would compare with our present results. Especially since the \textit{outliers} present a steep power index in the radio/X-ray plane at high luminosity, similar to the one observed in neutron star binaries. This is devoted to a forthcoming paper.

\begin{acknowledgements}
The authors acknowledge funding support from CNES and the French PNHE.
\end{acknowledgements}

\bibliographystyle{aa} 

\begin{appendix} 
\section{Outburst \#4 radio interpolation} \label{sec:a_interp}

\begin{figure}[H]
	\includegraphics[width=\columnwidth]{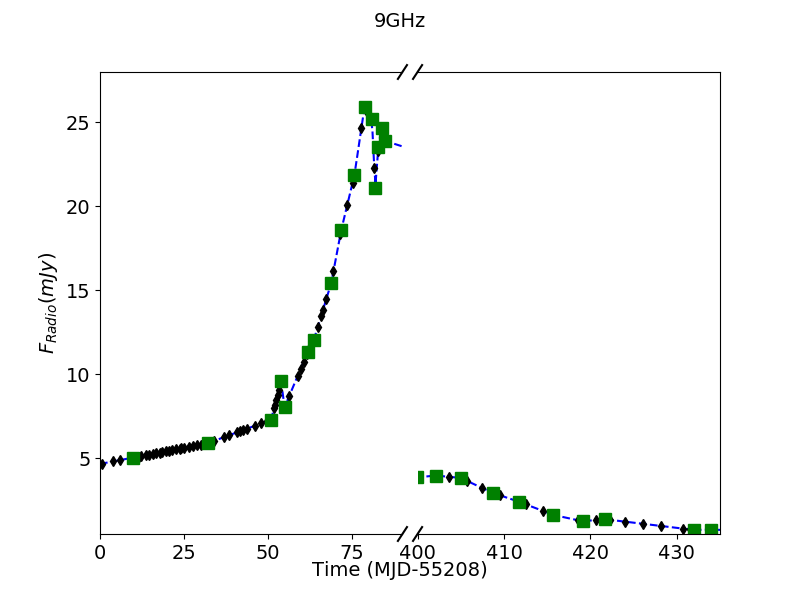}
    \caption{Radio light curve of outburst \#4 during its Hard states - between MJD 55208 and 55293 and between MJD 55608 and 55646. The green square are the observed radio data. The blue dashed line the linearly interpolated function. And the black diamonds are the interpolated radio fluxes at the date of the X-ray observations.}
    \label{fig:Radio_interp}
\end{figure}

With 80 almost daily X-ray observations and 24 radio fluxes measurements, the radio/X-ray survey of the hard states of the 2010 (\#4) outburst of GX339-4 is the best in the available archive of RXTE. Considering the steady jet expected during the hard states of an outburst and the evenly spread radio survey, we opted to interpolate linearly the radio light-curve as to obtain a radio flux for each X-ray observations. In Fig. \ref{fig:Radio_interp}, we plotted the radio light-curve. The green squares represent the radio observations. The dashed blue line represent the linear interpolation of the radio light-curve between the radio observations. The black diamonds represent the date at which X-ray observations were taken. 

\section{Fitting the decaying phase of outbursts \#4} \label{sec:b_decay2010}

\begin{figure*}[]
	\includegraphics[width=\linewidth]{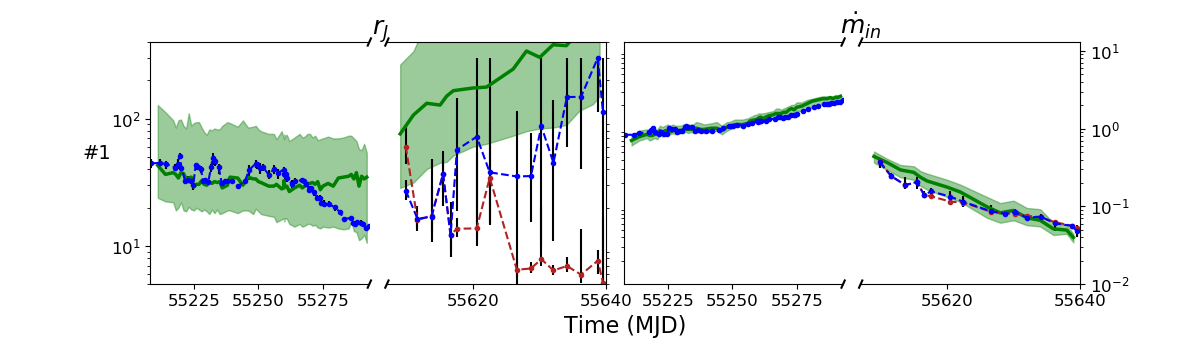}
    \caption{Results of the fitting procedure for outburst \#4. On the left side, the transition radius \rJ between the JED and the SAD. On the right side, the mass accretion rate \mdot. Each side is divided horizontally between the rising and decaying phase of each outburst. The green solid line represent the results from M19 and M20, and the green region where their minimization function varies by less than 10\% with respect to its minimum. The blue dashed line shows the results presented in Fig. \ref{Fig:Fits_results_norad} and the black vertical bar the associated 90\% confidence range. The red dashed line represent the initial fitting results for the decaying phase of outburst \#4.}
    \label{fig:Fits_2010}
\end{figure*}

In Fig. \ref{fig:Fits_2010}, we plotted the evolutions of \rJ and \mdot during outburst \#4. The blue dashed line are the final results presented in Fig. \ref{Fig:Fits_results_norad} used for the study. The red dashed line represent the initial results obtained by our fitting procedure. In this appendix, we explain why we rejected the results of this fitting procedure and how we obtained the final results for the decaying phase of outburst \#4 using a maximum likely-hood method.

\begin{figure}[h!]
	\includegraphics[width=\columnwidth]{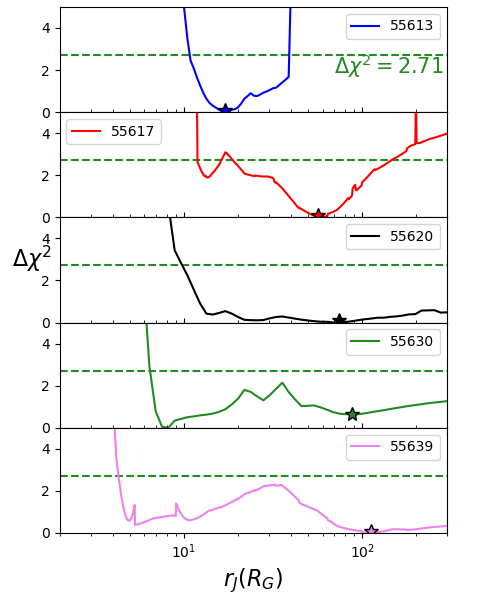}
    \caption{Evolution of the $\Delta\chi^2=(\chi^2(r_J)-\chi^2_{min})$ with $r_J$ for a few observations from the decaying phase of outburst \#4. From top to bottom in chronological order: MJD 55613, 55617, 55620, 55630, 55639. The 90\% confidence threshold $\Delta\chi^2=2.71$ is drawn as dashed green horizontal line. We highlighted with a star the solutions with higher values of $r_J$ we plotted in red in Fig. \ref{Fig:Fits_results_norad} and Fig. \ref{fig:Fits_2010}}
    \label{fig:steppar_rj}
\end{figure}

\begin{figure}[h!]
	\includegraphics[width=\columnwidth]{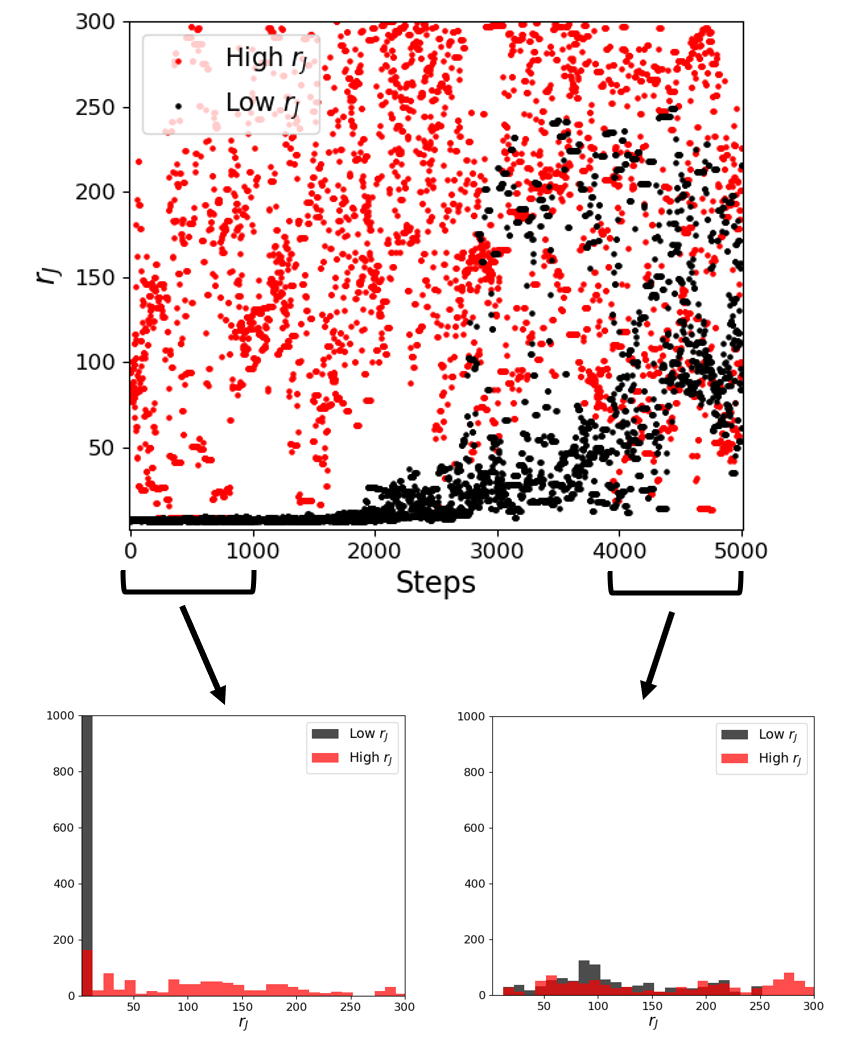}
    \caption{\textbf{Top panel:} MCMC procedure for observation 55630 (see the $\chi^2$ parameter space in Fig. \ref{fig:steppar_rj} in green) starting from two different initial priors. In black starting with a lower $r_J$ prior, in red starting with a higher $r_J$ prior. \textbf{At the bottom:} histograms of $r_J$ during the first 1000 steps (left) and the last 1000 steps (right). Both procedure end up exploring the higher $r_J$ solution only.}
    \label{fig:MCMC}
\end{figure}

The main disagreement of the fitting procedure with M19 was observed in the decaying phase of outburst \#4. The fits yielded quite small values of \rJ ($\sim 5 R_G$) and a decrease of the transition radius with time (Fig. \ref{fig:Fits_2010}, red), which is not expected in the JED-SAD paradigm and in contradiction with the evolution observed by M19, although their constraints are quite large. We report in Fig \ref{fig:steppar_rj} the evolution of the $\chi^2$ statistic along the parameter space of \rJ for a few observations of this decaying phase. These observations show a non-trivial $\chi^2$ space with multiple local minimums. Most of them presented either a better fit or a statistically equivalent fitting solution at higher $r_J$. For some reason, Xspec did not find these solutions even during the error calculations. In Fig. \ref{fig:MCMC} we represent MCMC tests for one of these observations (MJD 55630), starting from the higher (in red) and lower (in black) $r_J$ solutions. In both cases, the procedure converged to a reasonable solution within a few steps (< 5). When starting with the lower $r_J$ solution, the procedure first explores the lower $r_J$ values only. It is only after $\sim$ 2500 steps that it explores the higher $r_J$ solutions. And in the last 1000 steps, both procedure explores the higher $r_J$ solution only.
These higher $r_J$ solutions were further motivated by the fact that high values of \rJ naturally appear in outburst \#2 and \#3 (see Fig \ref{Fig:Fits_results_norad}). For these different reasons, we set $r_J$ to this solution for each observation of the decaying phase of outburst \#4. 

A similar process could have been applied to the decaying phase of the other outbursts, however the lack of radio flux where similar problems were encountered made this issue non critical for the rest of the study.


\newpage

\section{Effects of the systematic error} \label{sec:c_systematics}

In this section we show the effects of adding systematic errors to our radio fluxes. The reasoning behind this addition is that we accept a 10\% error on the reproduction of the radio fluxes. As the $\chi^2$ statistic gives more weight to the observations with small error, without systematic errors added, the fit is most of the time driven by a few fluxes and does not match our quality criterion. Our goal is to reproduce all the radio fluxes within a 10-20\% error margin. Adding systematic error to all fluxes increases the relative weight of the observations with higher uncertainty compared to the ones with low flux error.

\begin{figure}[H]
	\includegraphics[width=\columnwidth]{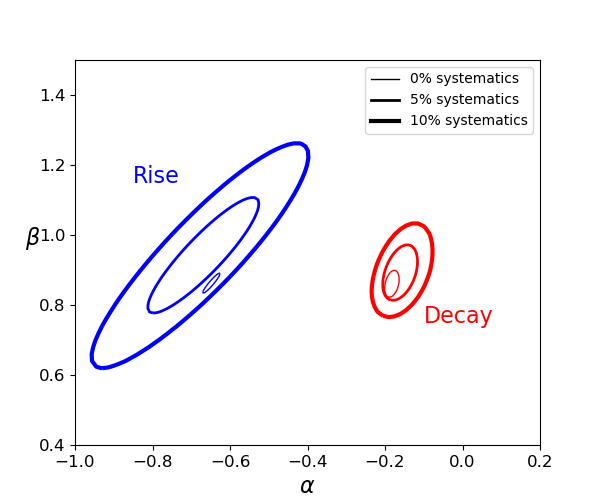}
    \caption{90\% Confidence contours $\beta$-$\alpha$ for different values of systematic error added to the radio fluxes (0\%, 5\% and 10\%). The thicker the line, the larger the systematic error added. In blue, the contours for all the rising phase observations. In red, the contours for all the decaying phase observations.}
    \label{fig:contour_syst_effects}
\end{figure}

However this has a major impact on the size of the confidence contour $\beta$-$\alpha$ that we obtain. Indeed these contour are directly depending on the $\chi^2$ plane of our parameter space. In Fig. \ref{fig:contour_syst_effects}, we plotted the confidence contour for a 90\% confidence level ($\Delta\chi^2=4.61$) for all the rising phase observations (in blue) and for all the decaying phase observations (in red) and for different values of systematic errors added to the radio fluxes (0\%, 5\% and 10\% depending on the thickness of the line). As expected, the larger the systematic error added, the larger the contour plot.
But even with 10\% systematic error, the rising phase solution and decaying phase solution are inconsistent.

\section{Impact of the other JED-SAD parameters} \label{sec:d_parameters}

\begin{figure*}[]
	\includegraphics[width=\linewidth]{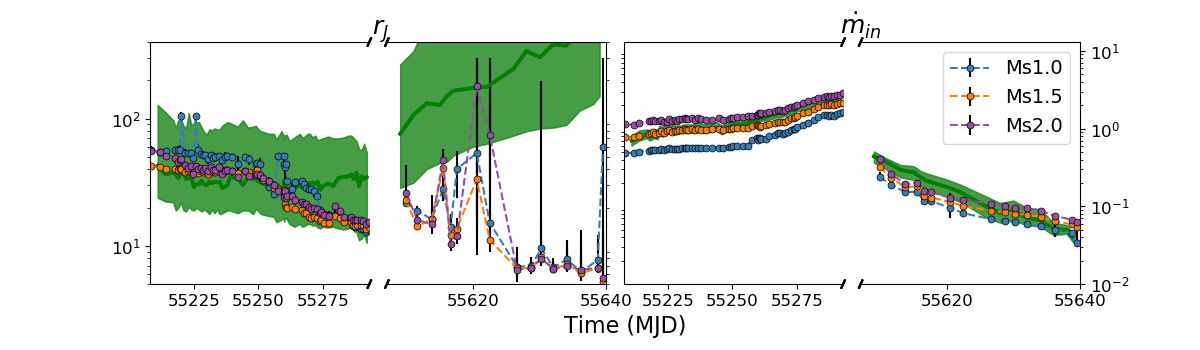}
    \caption{Results of the fitting procedure for outburst \#4. On the left side, the transition radius \rJ between the JED and the SAD. On the right side, the mass accretion rate \mdot. Each side is divided horizontally between the rising and decaying phase of each outburst. The green solid line represent the results from M19, and the green region where their minimization function varies by less than 10\% with respect to its minimum. The blue, orange and purple dashed lines show the results of the automatic fitting procedure using different values of $m_s$ (1.0, 1.5 and 2.0 respectively).}
    \label{fig:ms_2010}
\end{figure*}

\begin{figure*}[]
	\includegraphics[width=\linewidth]{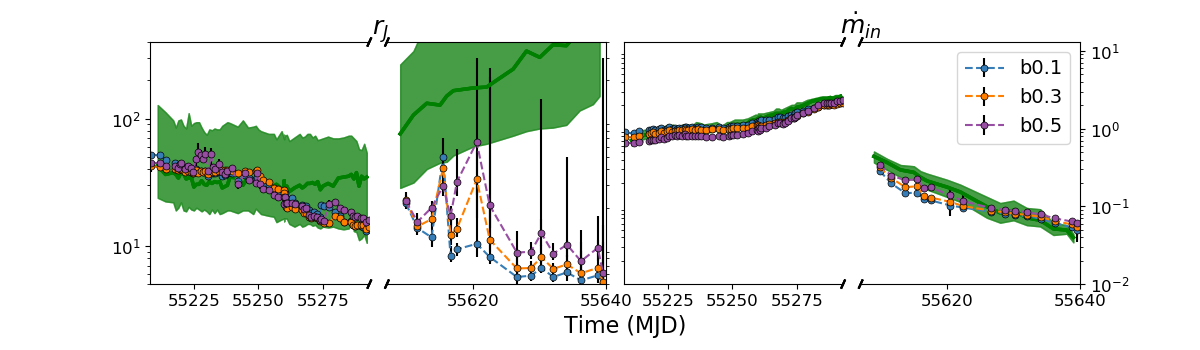}
    \caption{Results of the fitting procedure for outburst \#4. On the left side, the transition radius \rJ between the JED and the SAD. On the right side, the mass accretion rate \mdot. Each side is divided horizontally between the rising and decaying phase of each outburst. The green solid line represent the results from M19 and M20, and the green region where their minimization function varies by less than 10\% with respect to its minimum. The blue, orange and purple dashed lines show the results of the automatic fitting procedure using different values of $b$ (0.1, 0.3 and 0.5 respectively).}
    \label{fig:b_2010}
\end{figure*}

\begin{figure}[H]
	\includegraphics[width=\columnwidth]{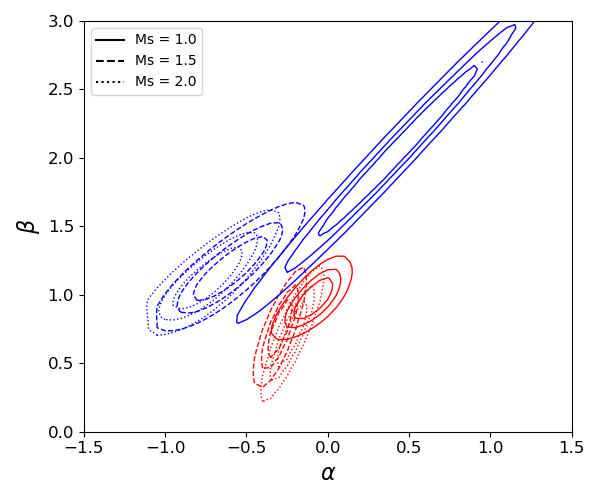}
    \caption{Confidence contours $\beta$-$\alpha$ for different values of $m_s$ used in the fitting procedure of outburst \#4. 1.0 in full line, 1.5 in dashed line and 2.0 in dotted line. In all cases, the rising phase contour and decaying phase contour are inconsistent. }
    \label{fig:contours_diff_ms}
\end{figure}

In this study we set the values of the parameters $m_s$, $p$ and $b$ to the one used by M19. \cite{marcel2018bunified} already performed a study of the parameter space and converged on these values. However, as the main spectral impact of $m_s$ and $b$ is a variation of the high-energy cutoff which is not visible inside of the \textit{RXTE}/PCA energy range, we are faced with unconstrained parameters. In this paragraph, we reproduce the study using other set values of $m_s$ and $b$. In Fig. \ref{fig:ms_2010} and \ref{fig:b_2010}, we plot the fitting results obtained for different values of $m_s$ and $b$. The global trend of the parameters stays similar, the values of $r_J$ and $\dot{m}_{in}$ are however slightly different. The biggest difference are observed with the different values of $m_s$. In Fig. \ref{fig:contours_diff_ms}, we plot the corresponding contour regions $\alpha$-$\beta$ for the different values of $m_s$. For the decaying phase observations, we use the procedure described in Appendix \ref{sec:b_decay2010} to obtain the transition radius $r_J$ for 8 quasi-simultaneous observations used for the contour plots. The contours obtained for different values of $m_s$ are not always consistent together. However, in each case, the rising phase and decaying phase contours are inconsistent. This implies that we always have two different behavior for the radio between the rising and decaying phase regardless of the values of $m_s$ and $b$ that we use in the fits.

\end{appendix}

\end{document}